\documentclass[12pt]{iopart}
\usepackage[switch]{lineno}
\expandafter\let\csname equation*\endcsname\relax
\expandafter\let\csname endequation*\endcsname\relax
\usepackage{amsmath} 
\usepackage{bm}
\usepackage{graphicx}
\usepackage{booktabs}
\usepackage{diagbox}
\usepackage{caption}
\usepackage{graphicx}
\usepackage{float}
\usepackage{subfig}
\usepackage{hyperref}
\usepackage{multirow}
\usepackage{xcolor}%

\begin{document}

\title{Output Prediction of Quantum Circuits based on Graph Neural Networks}

\author{Yuxiang Liu$^{1, 3, 4,\dag}$, Fanxu Meng$^{2,\dag}$, Lu Wang$^{4, 5}$, Yi Hu$^{1, 4}$, Zaichen Zhang$^{1, 3, 4,*}$ and Xutao Yu$^{3, 4, 5,*}$}

\address{$^1$ National Mobile Communications Research Laboratory, Southeast University, Nanjing, 210096, China}
\address{$^2$ College of Artificial Intelligence, Nanjing Tech University, Nanjing, 211800, China}
\address{$^3$ Purple Mountain Lab, Nanjing, 211111, China}
\address{$^4$ Frontiers Science Center for Mobile Information Communication and Security, Southeast University, Nanjing, 210096, China}
\address{$^5$ State Key Laboratory of Millimeter Waves, Southeast University, Nanjing, 210096, China}
\address{$*$ Corresponding authors}
\address{$\dag$ These authors contributed equally to this work.}

\ead{zczhang@seu.edu.cn, yuxutao@seu.edu.cn}

\begin{abstract}
The output prediction of quantum circuits is a formidably challenging task imperative in developing quantum devices. Motivated by the natural graph representation of quantum circuits, this paper proposes a Graph Neural Networks (GNNs)-based framework to predict the output expectation values of quantum circuits under noisy and noiseless conditions and compare the performance of different parameterized quantum circuits (PQCs). We construct datasets under noisy and noiseless conditions using a non-parameterized quantum gate set to predict circuit expectation values. The node feature vectors for GNNs are specifically designed to include noise information. In our simulations, we compare the prediction performance of GNNs in both noisy and noiseless conditions against Convolutional Neural Networks (CNNs) on the same dataset and their qubit scalability. GNNs demonstrate superior prediction accuracy across diverse conditions. Subsequently, we utilize the parameterized quantum gate set to construct noisy PQCs and compute the ground state energy of hydrogen molecules using the Variational Quantum Eigensolver (VQE). We propose two schemes: the Indirect Comparison scheme, which involves directly predicting the ground state energy and subsequently comparing circuit performances, and the Direct Comparison scheme, which directly predicts the relative performance of the two circuits. Simulation results indicate that the Direct Comparison scheme significantly outperforms the Indirect Comparison scheme by an average of 36.2\% on the same dataset, providing a new and effective perspective for using GNNs to predict the overall properties of PQCs, specifically by focusing on their performance differences.

\end{abstract}

%
\vspace{2pc}

\section{Introduction}
Quantum computing, as a new computational paradigm, has the potential to address problems beyond the capabilities of classical computing with greater efficiency and higher speed. It has demonstrated exponential or polynomial advantages in various domains, including combinatorial optimization \cite{12,13,14}, large-scale communication systems \cite{60,61,72,73}, molecular dynamics \cite{27,35}, quantum chemistry\cite{69,70,71}, and machine learning \cite{3,7,16,21,29,28,68}. Driven by breakthroughs in physical implementation technologies, quantum hardware has undergone rapid development over the past two decades, with numerous quantum computing systems now available \cite{15,19,23,62}. Despite its promising prospects, quantum computing must navigate an extended era of Noisy Intermediate-Scale Quantum (NISQ) devices before entering the fault-tolerant quantum computing era \cite{cheng2023noisy,36,74}. During this stage, the limitations imposed by coherence times and quantum gate errors pose significant bottlenecks to achieving quantum advantage. Quantum devices currently encounter limitations in resource accessibility, such as runtime and qubit count, which do not match the convenience provided by classical computers. Additionally, classical simulation methods often suffer from inefficiencies. Consequently, some advancements must proceed without adequate benchmark data, significantly impeding the engineering of quantum devices. Against this backdrop, accurately predicting the output of quantum circuits is not only a theoretical challenge but also a crucial step toward optimizing quantum algorithms and enhancing hardware performance. However, due to the complex noise characteristics of quantum circuits and the computational challenges that grow with circuit size, existing classical simulation methods often struggle to provide efficient solutions.

Notably, as it can provide low-cost predictions of specific aspects of quantum circuit outputs once the model is trained, machine learning has recently offered an attractive alternative to direct classical simulation. For instance, in \cite{cantori2023supervised}, a meticulously designed qubit-scalable Convolutional Neural Network (CNN) is employed to predict expectation values of circuits, delivering results that, under specific conditions, outperform those obtained from the freely available Noisy Intermediate-Scale Quantum (NISQ) devices. In \cite{63}, neural networks based on simple Multilayer Perceptrons (MLPs) are used to predict kernel-target alignment (KTA), a cost-effective proxy for quantum kernel classification accuracy, facilitating the identification of high-accuracy quantum kernel circuits. Despite these advancements, CNN-based and MLP-based approaches face limitations in capturing the structural information of quantum circuits, particularly for large-scale systems or circuits with complex topologies. Graph-based models, on the other hand, naturally leverage the directed acyclic graph (DAG) representation of quantum circuits, making them a promising alternative for output prediction. For instance, in \cite{wang2022quest}, Graph Transformers are utilized to predict the Probability of Successful Trials (PST), which positively correlates with circuit fidelity, achieving remarkable performance. More recently, \cite{liao2024machine} proposed a general framework of machine learning for quantum error mitigation (ML-QEM), which employs various machine learning models including Linear Regression, Random Forest (RF), and Graph Neural Networks (GNNs) to regress the noisy expectation values obtained from quantum processors toward their approximated noiseless counterparts for error mitigation. Compared with traditional mitigation methods, ML-QEM can significantly reduce the overall cost of quantum error mitigation. Beyond these applications, estimating or predicting quantum circuit outputs still holds broad prospects. For example, circuit output predictors could be integrated into diffusion models, such as those discussed in \cite{64}, to guide the generation of high-performance circuits. Alternatively, these predictors could support Reinforcement Learning (RL) processes, as outlined in \cite{65}, by evaluating reward functions, thereby reducing resource consumption during training.

Based on the considerations outlined above, we aim to design a Graph Neural Networks (GNNs) predictor to predict the specific outputs of quantum circuits. Within this approach, the GNNs act as a surrogate model that enables large-scale and rapid performance prediction once trained, eliminating the need for repeated costly simulations or hardware executions. In addition to computational efficiency, GNNs also contribute to improving prediction accuracy, as they can effectively capture the intrinsic structural dependencies within quantum circuits. This advantage arises not only from their ability to represent circuit topology more faithfully but also from their capability to construct individual feature vectors for each node, allowing the embedding of various properties of quantum gates, such as gate type, gate error, and target qubits. In this work, we primarily predict the single-qubit expectation values, two-qubit expectation values, and the overall property of quantum circuits.

\begin{figure}[htbp]
	\centering
	\includegraphics[width=0.95 \textwidth]{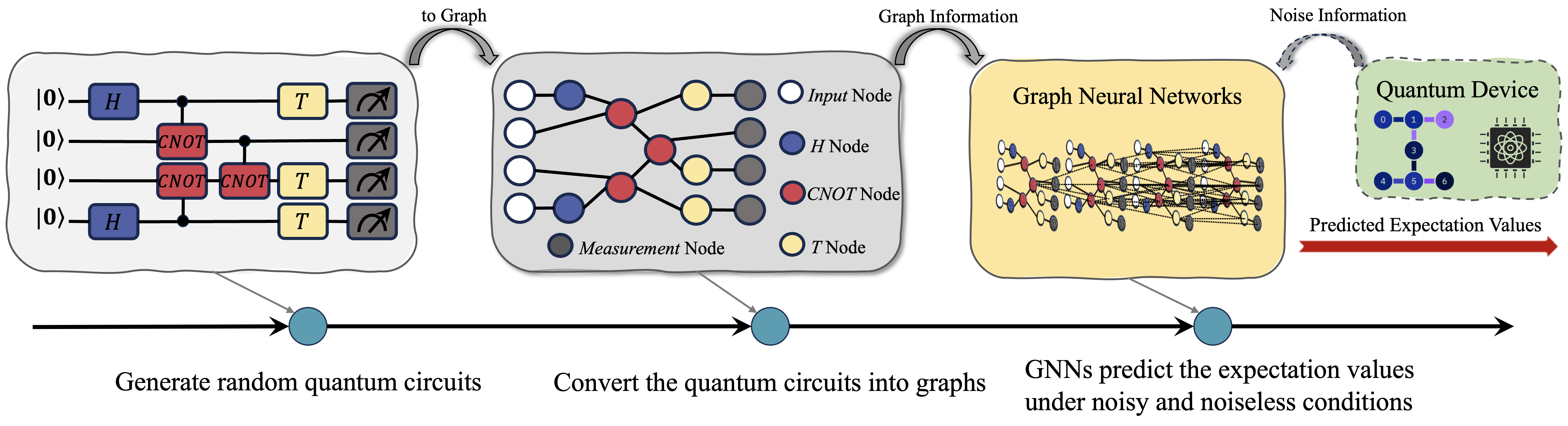}
	\captionsetup{justification=justified, singlelinecheck=false}
	\caption{The framework for expectation values prediction. a) Generate random quantum circuits. b) Transform the random quantum circuits into graph structures. c) Incorporate noise information into the graph nodes of the quantum circuits, train GNNs, and predict the expectation values of quantum circuits.} \label{fig3_1a}
\end{figure}
To begin with, we propose a GNN-based framework for predicting expectation values of quantum circuits, as illustrated in figure \ref{fig3_1a}. These limited expectation values are sufficient for circuits that require only one or two possible output bitstrings, such as the Bernstein–Vazirani (BV) algorithm \cite{66} and the Deutsch–Jozsa algorithm \cite{deutsch1992rapid}. Utilizing a parameter-free quantum gate set, we randomly generate quantum circuits under both noiseless and noisy conditions. Classical simulations are conducted using the Qiskit library \cite{67} to obtain single-qubit and two-qubit expectation values for our dataset. During the transformation of random circuits into graphs, we design feature vectors for each node to encapsulate properties such as gate type, target qubits, and noise information. Our experiments demonstrate that the GNNs achieve strong predictive performance on datasets with a relatively small circuit search space. For a fair comparison with the well-designed and representative convolutional neural networks (CNNs)-based method \cite{cantori2023supervised}, which possesses strong structured feature extraction capability, our GNNs-based method demonstrates superior performance, greater flexibility in handling input circuit structures, and improved scalability.

Furthermore, we extend our GNN-based prediction framework in two significant ways. First, we incorporate parameterized quantum gates, expanding its applicability to parameterized quantum circuits (PQCs). Second, instead of solely predicting expectation values, we shift our focus to evaluating the overall property of quantum circuits, specifically predicting the ground-state energy of the H$_2$ molecule encoded on 4 qubits using a standard VQE workflow. In parallel, recent studies have employed deep neural networks to predict VQE circuit parameters directly from molecular geometries, thereby bypassing expensive variational optimization \cite{tao2022exploring, ghosh2023deep}. Such ML-assisted VQE schemes have demonstrated accurate ground-state energy predictions, including for systems beyond H$_2$ (e.g., LiH and BeH$_2$), reinforcing the relevance of ML for PQC-based quantum chemistry. To compare the performance of PQCs, we propose two schemes: Direct Comparison, which directly predicts the relative performance difference between circuits, and Indirect Comparison, which involves predicting absolute ground state energies before comparison. Our simulation results demonstrate that the Direct Comparison scheme significantly outperforms the Indirect Comparison scheme, offering an average improvement of 36.2\% on the same dataset. These findings highlight the potential of GNNs in efficiently evaluating PQC performance, providing a novel and effective perspective for analyzing quantum circuits.

The rest of this paper is organized as follows: In Section \ref{sec2}, we introduce the background knowledge about quantum basics, quantum errors, and Graph Neural Networks. The frameworks for using GNNs to predict expectation values and compare circuit performance are detailed in Section \ref{sec3}. Section \ref{sec4} analyzes the prediction results and extrapolation capabilities of the trained GNNs, comparing them with CNNs-based methods. It also presents simulations and analyses of both direct and indirect comparison schemes. Finally, Section \ref{sec5} reports our conclusions and an outlook on future perspectives.

\section{Background}
\label{sec2}
\subsection{Quantum basics}
Quantum bits, or qubits, unlike classical bits, can exist in a linear superposition of the two basis states, 0 and 1, satisfying $\left | \psi  \right \rangle =\alpha \left | 0  \right \rangle +\beta \left | 1  \right \rangle $, where $\left | \alpha  \right | ^{2} + \left | \beta   \right | ^{2} = 1$. This superposition property allows an $n$-qubit system to represent a linear superposition of $2^{n}$ basis states. Quantum gates are used to perform computations on quantum systems by transforming quantum states from $\left | \psi    \right \rangle $ to $\left | \psi^{\prime }      \right \rangle $. Each quantum gate corresponds to a unitary matrix \cite{nielsen2010quantum}, and common single-qubit gates include Hadamard gate ($H$), phase gate ($S$), $\pi /8$ gate ($T$) and Pauli rotation gates ($R_{x} $, $R_{y} $, $R_{z} $), which can respectively be denoted as the following unitary matrices,
\begin{align}
T=\begin{pmatrix}
  1&0 \\
  0&e^{i \frac{\pi }{4} } 
\end{pmatrix}
&&
S=\begin{pmatrix}
  1&0 \\
  0&i 
\end{pmatrix}
&&H =\frac{1}{\sqrt{2} } \begin{pmatrix}
  1& 1\\
  1&-1
\end{pmatrix}
\end{align}

\begin{align}
R_{x} \left ( \theta  \right ) =\begin{pmatrix}
 \cos\frac{\theta}{2}  & -i\sin\frac{\theta}{2} \\
  -i\sin\frac{\theta}{2} & \cos\frac{\theta}{2}
\end{pmatrix}
&&
R_{y} \left ( \beta  \right ) =\begin{pmatrix}
 \cos\frac{\beta}{2}  & -\sin\frac{\beta}{2} \\
  \sin\frac{\beta}{2} & \cos\frac{\beta}{2}
\end{pmatrix}
&&
R_{z} \left ( \gamma   \right ) =\begin{pmatrix}
 e^{-i\frac{\gamma }{2}}  & 0 \\
  0 & e^{i\frac{\gamma }{2}}
\end{pmatrix}
\end{align}
The common two-qubit gate is the CNOT gate as follows
\begin{align}
  CNOT=\begin{pmatrix}
 1 & 0 & 0 & 0\\
 0 & 1 & 0 & 0\\
 0 & 0 & 0 & 1\\
 0 & 0 & 1 &0
\end{pmatrix}
\end{align}

\subsection{Quantum errors}
Quantum errors pose one of the major challenges for quantum computing in the NISQ era. On real quantum devices, errors arise from interactions between qubits and their environment, control errors, and environmental disturbances \cite{4,25,33}. As quantum devices operate, qubits undergo coherence errors over time, while quantum gates introduce operational errors, such as coherent or stochastic errors. These errors significantly disrupt the functionality of quantum circuits and hinder their further optimization. To address these challenges, various noise mitigation techniques have been proposed to reduce the negative effects of quantum errors \cite{17,26,29,37,50,51,52}.

\subsection{Graph Neural Networks}
Graph Neural Networks (GNNs) have emerged as a transformative approach for processing and learning from graph-structured data, demonstrating exceptional capabilities in understanding the complex relational structures inherent in such data \cite{53}. By leveraging a message-passing mechanism that aggregates information from neighboring nodes while capturing the relational structure of the graph, GNNs provide a solid foundation for accurate predictions and informed decision-making \cite{54}. In recent years, various GNN variants, such as Graph Convolutional Networks (GCNs), Graph Attention Networks (GATs), and GraphSAGE, have achieved groundbreaking performance across numerous deep learning tasks \cite{55,56,57,58,59}. Quantum circuits, with their inherent connectivity relationships, can be naturally represented as graph structures, offering an intuitive pathway for applying GNNs in quantum computing.

\section{Methods} 
\label{sec3}
In this section, we use GNNs to predict the outputs of two different classes of quantum circuits, where the universal gate sets that constitute the two classes of quantum circuits are different. In Section 3.1, we use GNNs to predict the expectation values of single-qubit and two-qubit under noisy and noiseless conditions. In Section 3.2, we extend the usage scenario of GNNs in Section 3.1 to realize the performance comparison of quantum circuits with different structures.

\subsection{GNNs for expectation values prediction}
\label{sec3.1}
Due to the limited quantum resources of the NISQ devices, it is a meaningful thing to predict the output of the circuit in advance before submitting it for execution. If the predicted output of the circuit is far below a threshold, running this circuit on a real quantum computation will not produce the desired results.
To accurately predict circuit output, especially in the presence of noise, we propose a framework for predicting quantum circuits with GNNs, as shown in figure \ref{fig3_1a}. This is the data-driven approach, and intuitively, estimating expectation values does not require computing the complete density matrix exactly.
Therefore, data-driven approaches present the opportunity to provide sufficiently accurate predictions at a lower computational cost.
The framework for predicting expectation values is divided into three primary stages. The first stage is to generate random quantum circuits with the assistance of a universal gate set and measure the circuit expectation values, which are used as training data for the GNNs. The second stage is to transform the generated random circuits into a graph structure and formulate feature vectors for each node that can contain device noise information. In the third stage, the graph is processed to predict circuit expectation values using GNNs.

This work has practical value as the framework realizes accurate prediction of the output expectation values of the circuits.
When only one-bit string has non-zero measurement probability in the circuit, the single-qubit expectation values are sufficient to find it. When combined with two-qubit expectation values, it is enough to identify the output bit strings when only two of them have non-zero probability \cite{cantori2023supervised}. Some relevant quantum algorithms, such as the Deutsch–Jozsa algorithm \cite{deutsch1992rapid}, the quantum counting algorithm \cite{nielsen2010quantum}, and the Grover algorithm with two searched items \cite{grover1996fast}, can assist in solving problems related to the prediction of the expectation values mentioned above. In the following, we will discuss the specific implementation details of the framework outlined in figure \ref{fig3_1a}.

\subsubsection{Random circuits generation}
\label{sec3.1.1}
\begin{figure}[thbp]
    \centering
    \includegraphics[width=0.96 \textwidth]{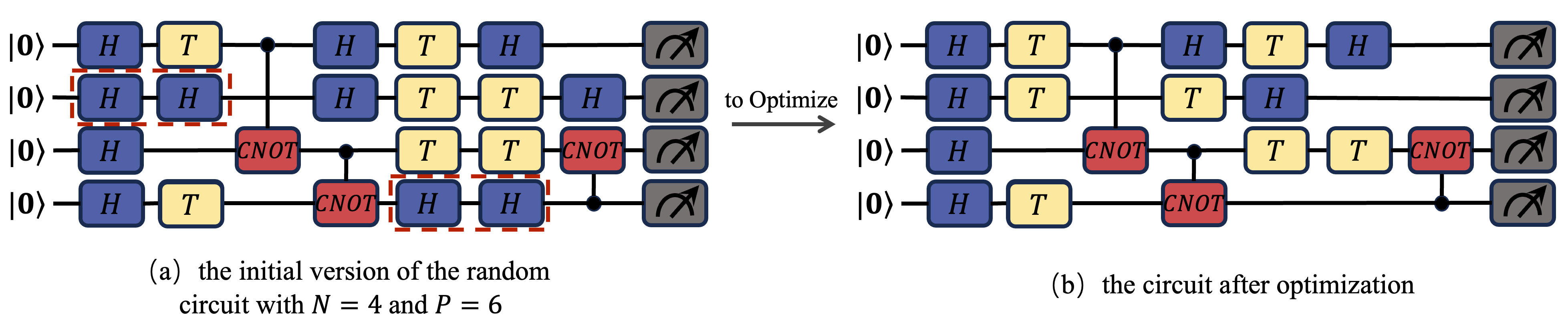}
    \captionsetup{justification=justified, singlelinecheck=false}
    \caption{Random circuits generation and circuits optimization.} \label{fig3_1b}
\end{figure}

\ 
\newline
\indent To realize the first stage in our framework, random quantum circuits are generated to provide a database for subsequent GNNs training. We select two single-qubit gates, namely the $\pi /8$ gate ($T$) and Hadamard gate ($H$), and one two-qubit gate CNOT, as the universal gate set to generate random circuits. The universal gate set consisting of parameter-free gates is denoted $S=\left \{ T, H, CNOT\right \} $. In the first step, gates are randomly selected from a universal gate set $S$ to create an initial version of the random circuit with $N$ qubits and $P$ layers of gates, which we define as the circuit depth. The integer $N$ corresponds to the number of qubits, and $P$ denotes the number of gates on each qubit, where two-qubit gates are considered as separate single-qubit operations acting on both qubits. To ensure circuits integrity, we set the initial state of all qubits to $\left | 0 \right \rangle$, and the circuits include measurement operations. After completing the circuit construction, an optimization strategy is used to eliminate duplicate gates to obtain circuits for later training. These optimization strategies include the equivalence of two consecutive H-gates to an Identity gate, the equivalence of eight consecutive T-gates to an Identity gate, and the elimination of two consecutive CNOTs, etc. The motivation for this step is to ensure that the circuits in our training dataset remain as compact as possible, so that the neural network can focus on learning meaningful structural differences rather than spending capacity on trivial equivalence transformations. The specific process of generating random circuits is illustrated in figure \ref{fig3_1b}. The quantum gates enclosed by the red dashed box in figure \ref{fig3_1b}(a) are the optimizable gates, and the optimized circuit is shown in figure \ref{fig3_1b}(b).
The number of randomly generated circuits in a single run, i.e., the training data size for the GNNs, is on the order of $10^4$, $N_{train} \simeq  10^4$.

\subsubsection{Graph construction}
\label{sec3.1.2}
\ 
\newline
\indent Initially, we employ directed acyclic graphs (DAGs) to depict the topology of quantum circuits. Each node corresponds to a qubit, quantum gate, or measurement, while edges represent the time-dependent sequence of various gates.
As depicted on the left side of figure \ref{fig3_1a}, perform the transformation from the circuit to graph structure, with particular attention to representing the two-qubit CNOT using a single graph node.

\subsubsection{Node features}
\label{sec3.1.3}
\begin{figure}[thbp]
    \centering
    \includegraphics[width= \textwidth]{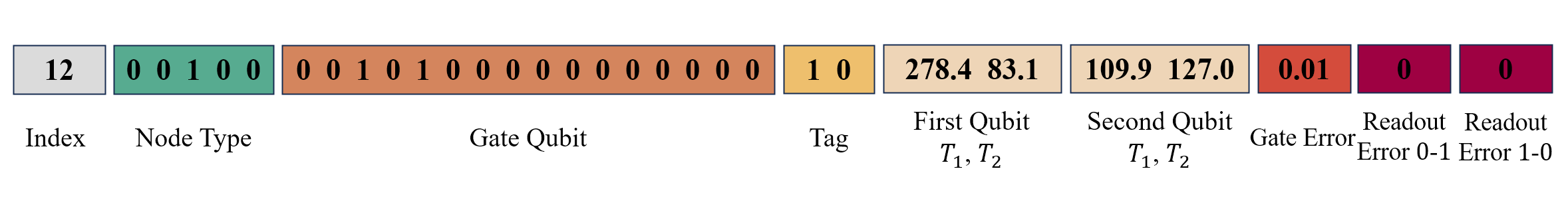}
    \captionsetup{justification=justified, singlelinecheck=false}
    \caption{Node feature vector.} \label{fig3_1c}
\end{figure}

\ 
\newline
\indent We construct a feature vector for each node in the graph to represent certain attributes of the node. The features include index, node type, gate qubit, tag, the relaxation time ($T_{1}$) and dephasing time ($T_{2}$) of the target qubits, gate error, and readout error, as shown in figure \ref{fig3_1c}.
The length of the node feature vector is 31. The first part is the index of node, which is used for node sorting and differentiation. The subsequent three parts are one-hot encoded representations of the node type, gate qubit, and tag. Here, the node type indicates the gate type: initial input qubit, measurement, CNOT, H, or T. For example, $00100$ signifies that the node represents a CNOT gate.
The gate qubit is used to label the index of the qubit affected by the quantum gate, and its length can be expanded according to practical needs. The tag is effective when the gate is a CNOT, indicating the control qubit and the target qubit of the CNOT.
One advantage of GNNs is their ability to consider hardware noise, primarily achieved through feature vectors. In this experiment, we utilize the last 7 numbers to characterize the noise properties of the hardware. The noise information is mainly referenced from IBM's quantum devices, such as IBM Perth, IBM Lagos, IBM Nairobi, etc. The first 4 numbers respectively represent the $T_1$ and $T_2$ values for the first and second qubits. The 5th number denotes the gate error for the gate, while the last two numbers indicate readout errors. If a node does not possess a specific noise characteristic, it is set to 0. For example, for a measurement node, $T_1$, $T_2$, and gate error are all set to 0.
For a node representing the CNOT, its complete feature vector is illustrated in figure \ref{fig3_1c}.

\subsubsection{Target values}
\label{sec3.1.4}
\ 
\newline
\indent After the circuit, the output state can be expressed as
\begin{equation}
\left | \bm{\varphi}   \right \rangle =U\left | 0 \right \rangle ^{\otimes N}
\end{equation}
Where $U$ denotes the unitary operator corresponding to the circuit, and the tensor product $\left | 0 \right \rangle ^{\otimes N}\equiv \left | 0  \right \rangle \otimes \dots \otimes \left | 0  \right \rangle $ is the initial state.
We define $Z_{i}$ as the Pauli operator $Z$ acting on qubit $i$, $i\in \left \{ 1,\dots ,N \right \}$ .
The single-qubit expectation values can be expressed as
\begin{equation}
\left \langle Z_{i} \right \rangle = \left \langle \bm{\varphi} | Z_{i} | \bm{\varphi}  \right \rangle 
\end{equation}
Since $Z_i = \left| 0 \right\rangle_i \left\langle 0 \right|_i - \left| 1 \right\rangle_i \left\langle 1 \right|_i$, the expectation value of $\left \langle Z_{i}  \right \rangle $ can thus be calculated as
\begin{equation}
\left \langle Z_{i}  \right \rangle =\left | \left \langle \bm{\varphi}  | 0  \right \rangle _{i}  \right | ^{2} -\left | \left \langle \bm{\varphi}  | 1  \right \rangle _{i}  \right | ^{2}
\end{equation}
For convenience, we constrain the single-qubit expectation values to be between 0 and 1, with an expectation value of 0 for 
$\left | 0  \right \rangle $ and an expectation value of 1 for $\left | 1  \right \rangle $. We scale the above formula accordingly:
\begin{equation}
z_{i}=1-\frac{\left \langle Z_{i} \right \rangle + 1}{2} 
\label{equation3_1c}
\end{equation}
The $z_{i}$ is the target value that denotes the expectation value of qubit $i$. Similar to the definition method for single-qubit expectation values, we define the scaling form for two-qubit expectation values:
\begin{equation}
z_{ij}=1-\frac{\left \langle Z_{i}Z_{j} \right \rangle + 1}{2}
\label{equation3_1d}
\end{equation}
Where $z_{ij}$, $i,j=1,2,\dots ,N$ represents the two-qubit expectation values.
Certainly, the description provided by the expectation values of single-qubit and two-qubit Pauli-Z measurements is limited.
However, these pieces of information are sufficient for certain specific algorithms, such as the Deutsch–Jozsa algorithm \cite{deutsch1992rapid}. For instance, in the Deutsch–Jozsa algorithm, the goal is to determine whether a Boolean function $f:\left \{ 0,1 \right \}^{\left ( N-1 \right ) }\longrightarrow   \left \{ 0,1 \right \} $ is constant or balanced. The algorithm measures the first $N-1$ qubits, and if the outcome corresponds to the bit string $00\cdots 0$  the function is constant; otherwise, it is balanced. In other words, predicting the expectation values of the $N-1$ single-qubit observables $\left \langle Z_{i}  \right \rangle $ (for $i=1,2,\cdots ,N-1$) is already sufficient to infer the result, that is, if $\left \langle Z_{i}  \right \rangle =1$ for all $i=1,2,\cdots ,N-1$ , the output bitstring must be $00\cdots 0$, indicating a constant function. Conversely, the function is balanced. This motivation underlies our exploration of expectation values prediction.

\subsubsection{Construction of GNNs}
\label{sec3.1.5}
\begin{figure}[thbp]
    \centering
    \includegraphics[width= \textwidth]{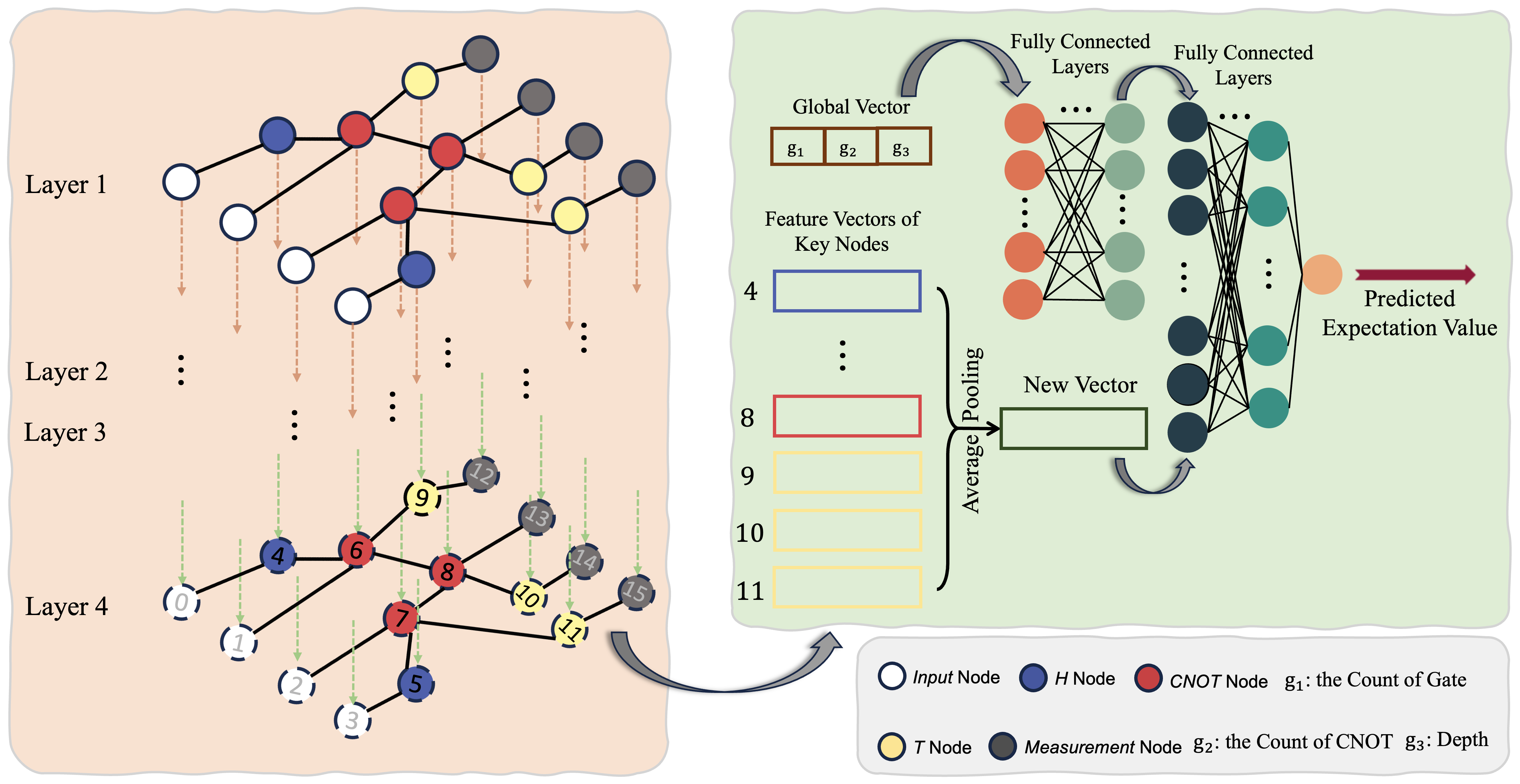}
    \captionsetup{justification=justified, singlelinecheck=false}
    \caption{Overview of the GNNs for the expectation value prediction.} \label{fig3_1d}
\end{figure}

\ 
\newline
\indent To process graphs with node features, we designed the four-layer Graph Neural Networks (GNNs), as illustrated in figure \ref{fig3_1d}, and incorporated an attention mechanism between each layer.
By leveraging the attention mechanism, the weights of each edge in the graph are dynamically determined, enabling the neural network to better capture information across the entire circuit. After feature extraction by the GNNs, features from key nodes (excluding input and measurement nodes) are selected and subjected to average pooling to obtain a 31-dimensional pooled vector. Meanwhile, the global circuit information, including the total number of gates, the number of CNOT gates, and the circuit depth, is fed into a two-layer fully connected network, where both the hidden and output layers have a dimensionality of 12. These global features are then concatenated with the pooled feature vector. Finally, this concatenated vector is processed through a three-layer fully connected network with hidden layer dimensions of 256, 128, and 64 to produce the final prediction result. The activation function between layers is implemented using the LeakyReLU function with a coefficient of 0.02.

\subsection{GNNs for circuit performance comparison}
\label{sec3.2}
In this chapter, we primarily expand upon the framework introduced in Section 3.1. Firstly, we replace the universal gate set for generating random quantum circuits, transitioning from the previous set $S=\left \{ T, H, CNOT\right \} $ to $S^{\ast } =\left \{ R_{x}\left ( \theta  \right ), R_{y}\left ( \beta  \right ), CNOT\right \} $.
The updated universal gate set allows for the generation of parameterized quantum circuits (PQCs), significantly broadening the applicability of quantum circuits. For example, this tool can be utilized in the search for optimal PQC structures. Secondly, we extend the values predicted by the GNNs, enabling them not only to predict the expectation values of qubits (including single-qubit and two-qubit expectation values) but also to predict the overall properties of a quantum circuit. Finally, we choose the Variational Quantum Eigensolver (VQE) algorithm as an extension application of our Section 3.1 architecture. We achieve this by using randomly generated PQCs and the optimized parameters to determine the ground state energy values of the molecule, which are then used as data to train the GNNs.
The ground state energy of the molecule is optimized through the following formula \cite{tilly2022variational} by adjusting the parameters of different circuits:
\begin{equation}
E_{VQE}=\min_{\bm{\theta}  }  \left \langle \bm{0}\left | U^{\dagger }\left ( \bm{\theta}  \right )H U \left ( \bm{\theta}  \right )    \right | \bm{0} \right \rangle
\end{equation}
where $H$ denotes the Hamiltonian operator corresponding to the lowest eigenvalue, and $U \left ( \bm{\theta}  \right ) $ denotes the parameterized ansatz, which also represents the PQC in the process of circuit design and implementation. 
By using the ground state energy values calculated through the Variational Quantum Eigensolver (VQE) as performance metrics for different circuits, and employing these energy values as data to train GNNs, we can predict the performance of randomly generated circuits. Specifically, we use the commonly adopted VQE optimizer COBYLA to optimize the circuit parameters. COBYLA is a gradient-free optimization method that does not require explicit computation of parameter gradients. The number of optimization iterations is set to 200, after which the optimized ground-state energy is obtained and used as the training target for the GNNs. It should be noted that the GNNs do not participate in the VQE optimization process. The VQE is solely employed to generate ground-state energy data for different circuits, which are used as labels for training the GNNs. It is worth noting that the predicted energy values are jointly affected by both the circuit structure and the optimization process. Although our model does not explicitly simulate the optimization process itself, it indirectly reflects the quality of circuit structures by learning the final energies of different circuits optimized under a fixed optimizer. This approach lays the foundation for selecting high-performance circuits. In the following two subsections, we will present this work. Section 3.2.1 utilizes the GNNs to predict the ground state energy of a molecule initially and compares the performance of PQCs based on the predicted ground state energy values. Section 3.2.2 builds upon the foundation of Section 3.2.1 by introducing an innovative approach: two different PQCs are simultaneously input into the GNN, which directly predicts the relative probability of their performance comparison.

\subsubsection{Indirectly comparing circuit performance}
\label{sec3.2.1}
\begin{figure}[thbp]
    \centering
    \includegraphics[width= 0.68 \textwidth]{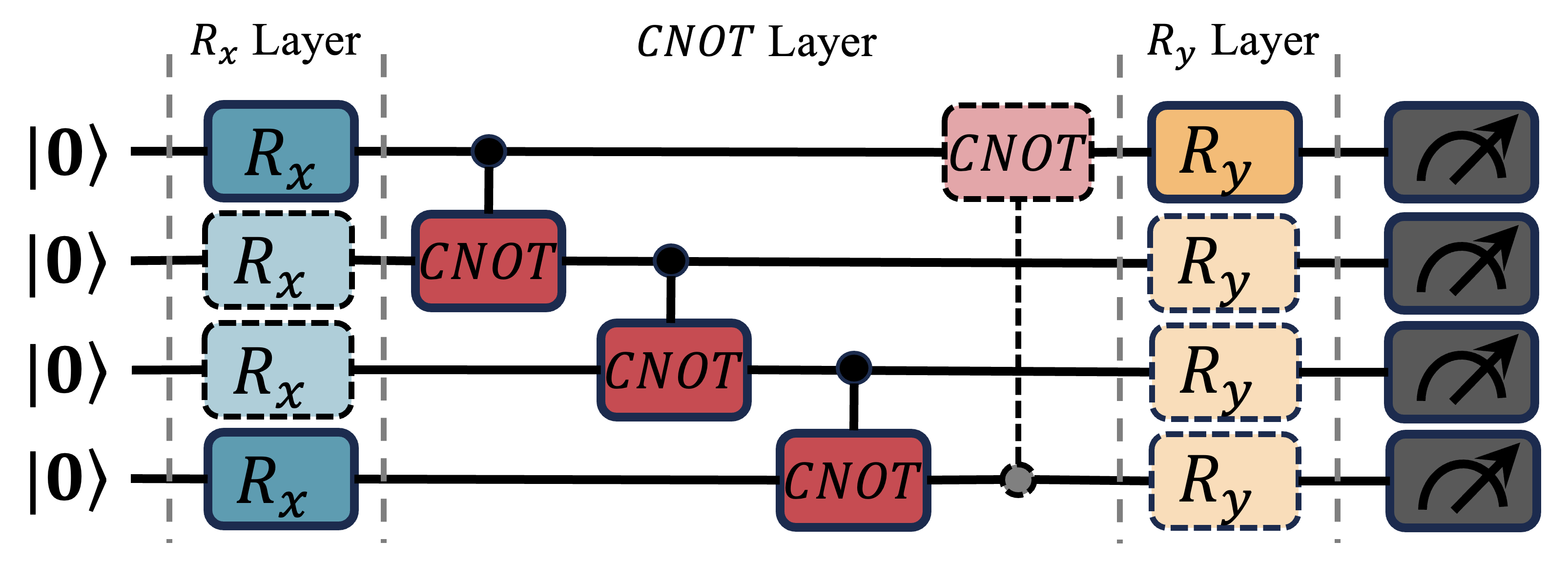}
    \captionsetup{justification=justified, singlelinecheck=false}
    \caption{Randomly generated single layer parameterized quantum circuit (PQC) with the quantum gates enclosed in dashed boxes representing those not selected.} \label{fig3_2a}
\end{figure}

\ 
\newline
\indent Differing from the method of generating random circuits in Section 3.1.1, we use $S^{\ast } =\left \{ R_{x}\left ( \theta  \right ), R_{y}\left ( \beta  \right ), CNOT\right \} $ as the universal gate set. We randomly generate PQCs with three layers, each layer consisting of a $R_{x}$ layer, a CNOT layer, and a $R_{y}$ layer. A single layer of the parameterized quantum circuit (PQC) is randomly generated as shown in figure \ref{fig3_2a}. To construct a three-layer PQC, the process shown in figure \ref{fig3_2a} is repeated three times, and the resulting three circuits are concatenated to form a complete quantum circuit.
For the $R_{x}$ layer, each qubit has $0.5$ probability of having a $R_{x}$ gate, and similarly for the $R_{y}$ layer. For the CNOT layer, there is $0.5$ probability of adding a CNOT gate between each adjacent pair of qubits, including the case where the first and last qubits are considered adjacent and one of the qubits is randomly selected as the control qubit. The generated PQCs are then subjected to noise, and after multiple iterations of classical optimization algorithms, we obtain approximate ground state energy values. These ground state energy values serve as the predicted outputs for training. The node features and the construction of GNNs are consistent with Section 3.1.

We directly employ GNNs to predict the ground state energy values corresponding to different PQCs. These predicted values are then used as a benchmark for assessing the relative performance of different circuits.

\subsubsection{Directly comparing circuit performance.}
\label{sec3.2.2}
\begin{figure}[thbp]
    \centering
    \includegraphics[width= 0.88 \textwidth]{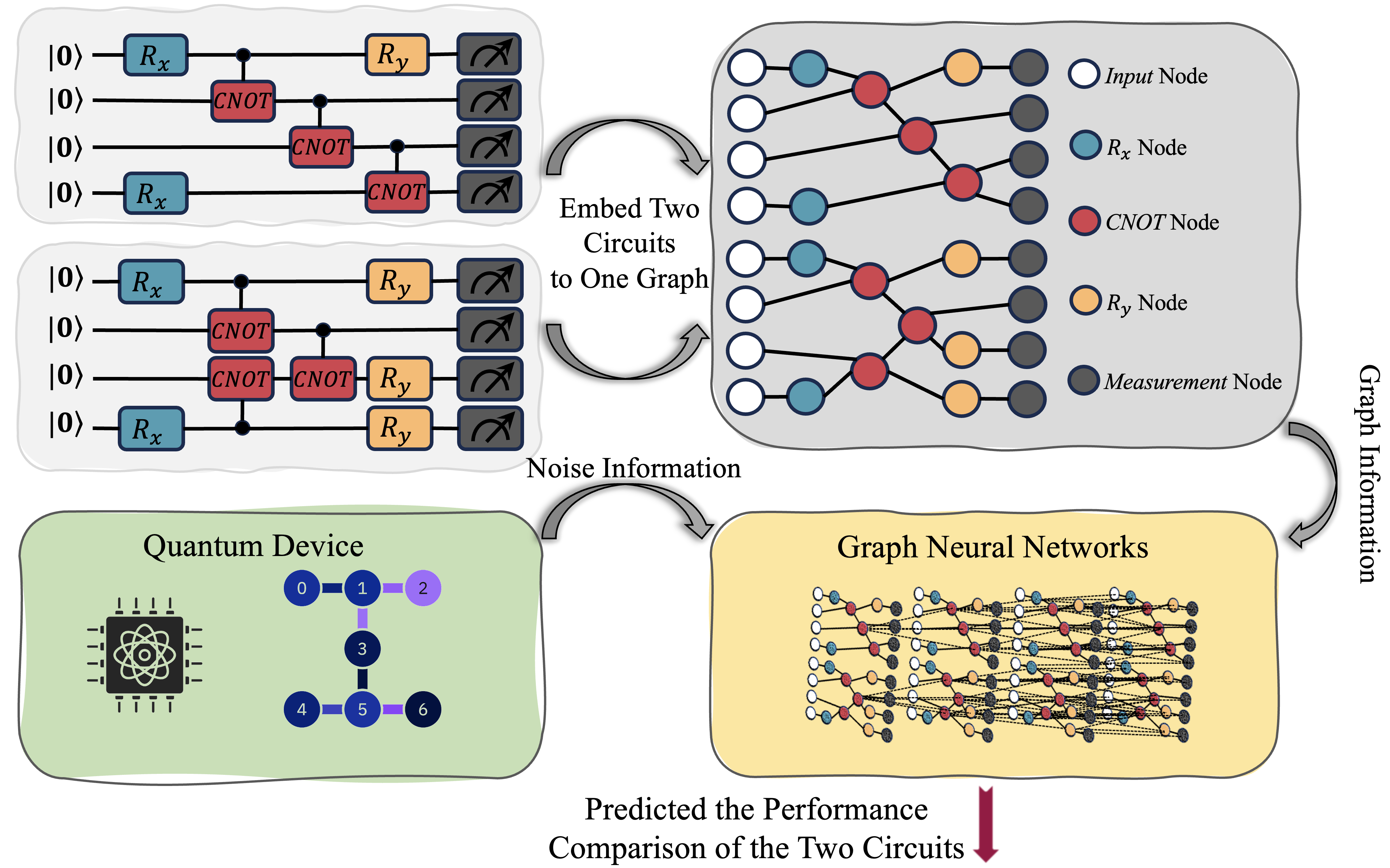}
    \captionsetup{justification=justified, singlelinecheck=false}
    \caption{The framework of circuit performance comparison prediction. Firstly, convert the two quantum circuits into one graph, where each node feature vector includes information about the noise. Input the graph into the GNNs to directly predict the performance comparison of the two circuits.} \label{fig3_2b}
\end{figure}

\ 
\newline
\indent Differing from Section \ref{sec3.2.1}, where circuit performance is indirectly compared by predicting the ground state energy values corresponding to the optimization parameters, we propose a new approach. In this new approach, we simultaneously input the two circuits that need to be compared into the GNNs. This allows the GNNs to learn the structural differences between the two circuits and directly predict the performance comparison between them. The specific framework for predicting circuit performance comparison using GNNs is illustrated in figure \ref{fig3_2b}.

During the graph construction process, two quantum circuits are merged into a single graph consisting of two disconnected subgraphs, as illustrated in the upper part of figure \ref{fig3_2b}. In the four-layer Graph Neural Networks (GNNs), each of the first three layers performs not only node-wise message aggregation but also average pooling on the key nodes (excluding input and measurement nodes) within each subgraph. The pooled feature vectors from the two subgraphs are then subtracted to obtain the difference features. Differently from the previous layers, the final layer also performs average pooling on the key nodes of both subgraphs but directly concatenates the two pooled feature vectors instead of taking their difference, aiming to preserve more individual information from each subgraph after aggregation. The difference vectors obtained from the first three layers and the concatenated feature vector from the final layer are then concatenated to form the output representation of the GNNs. Meanwhile, global information, including the differences in the total number of quantum gates, CNOT gates, and circuit depths between the two circuits, is first passed through two fully connected layers. The resulting output is then concatenated with the GNNs output to form a unified feature vector, which is subsequently fed into three fully connected layers to produce the final prediction. During the training process, the noise model used is derived from IBM’s quantum device, IBM Lagos. Furthermore, the output of the GNNs is the probability that the first circuit outperforms the second circuit, ranging from 0 to 1. To summarize the distinction between the two comparison schemes, in the indirect comparison scheme, the GNNs predict the ground-state energy obtained from VQE optimization for each circuit, whereas in the direct comparison scheme, the inputs are two circuit graphs, and the output represents the probability that the first circuit performs better than the second one. The labels are constructed based on the energy difference between the two circuits after Min–Max normalization.

\section{Evaluation} 
\label{sec4}
\subsection{Evaluation methodology}
\label{sec4.1}
\textbf{Model and Training Setups.} In the default setup, we use four-layer Graph Neural Networks. The embedding dimension is set to 31, corresponding to the feature vector dimension of 31. We employ two-head attention layers. By performing average pooling on key nodes, a 31-dimensional vector is generated as the aggregated feature for the circuit. If global features are incorporated into the GNN, we use two fully connected (FC) layers to expand the dimensions of the global features, with both the hidden and output dimensions set to 12. These global features are then concatenated with the pooled aggregated features. The concatenated features are further processed through three FC layers with hidden dimensions of 256, 128, and 64, respectively. The output dimension is set to 1 or 2, depending on the prediction target. We use the LeakyReLU activation function with a hyperparameter $\alpha =0.02$. During data preprocessing, Min-Max normalization is applied to the node features across the entire dataset. The model is trained for 200 epochs using the Adam optimizer with a constant learning rate of 0.01, a batch size of 512, and the mean squared error (MSE) loss function. Finally, during the evaluation phase, the coefficient of determination is used:
\begin{equation}
R^{2}=1-\frac{ {\textstyle \sum_{k=1}^{N_{test} } } {\textstyle \sum_{i=1}^{N}} \left ( y_{i}^{\left ( k \right ) } - \hat{y} _{i}^{\left ( k \right ) }\right ) ^{2}  }{ {\textstyle \sum_{k=1}^{N_{test} } } {\textstyle \sum_{i=1}^{N}} \left ( y_{i}^{\left ( k \right ) } - \bar{y}  \right ) ^{2} } 
\end{equation}
where $\hat{y} _{i}^{\left ( k \right ) }$ is the output prediction associated with the ground truth target value $y_{i}^{\left ( k \right ) }$, $\bar{y}$ is the average of the target values, $N_{test}$ is the number of circuits in the test set, and $N$ is the number of outputs. It is worth emphasizing that $R^{2}$ quantifies accuracy in relation to the intrinsic variance of the test data. This metric is suitable for fair comparisons across circuits of different scales, as circuits of varying sizes may exhibit a tendency for output values to cluster around or near the mean to varying degrees.
\newline
\textbf{Dataset Setup.} For the noisy and noiseless simulator datasets, we randomly generate 20000 random circuits for each case. In addition to noise information derived from real quantum devices (IBM Perth, IBM Lagos, IBM Nairobi, and IBM Jakarta), we also define a noise model, referred to as Simulated Noise, in which gate errors,  $T_{1}$,  $T_{2}$, and readout errors are fixed at predefined values. The detailed parameters of the noise model are provided in \ref{Appendix_A}. During the extrapolation process for qubits, we randomly generate 40000 random circuits as the dataset.

\subsection{Expectation values prediction}
\label{sec4.2}
To explore the advantages of using GNNs for circuit output prediction, we conduct extensive experiments. These experiments included predicting the single-qubit output expectation values in both noisy and noiseless conditions, predicting the two-qubit output expectation values under the same conditions, and evaluating the extrapolation capability of GNNs in predicting circuit outputs in both noisy and noiseless environments.
\subsubsection{Single-qubit expectation values}
\label{sec4.2.1}
\begin{figure}[thbp]
    \centering
    \includegraphics[width= 0.88 \textwidth]{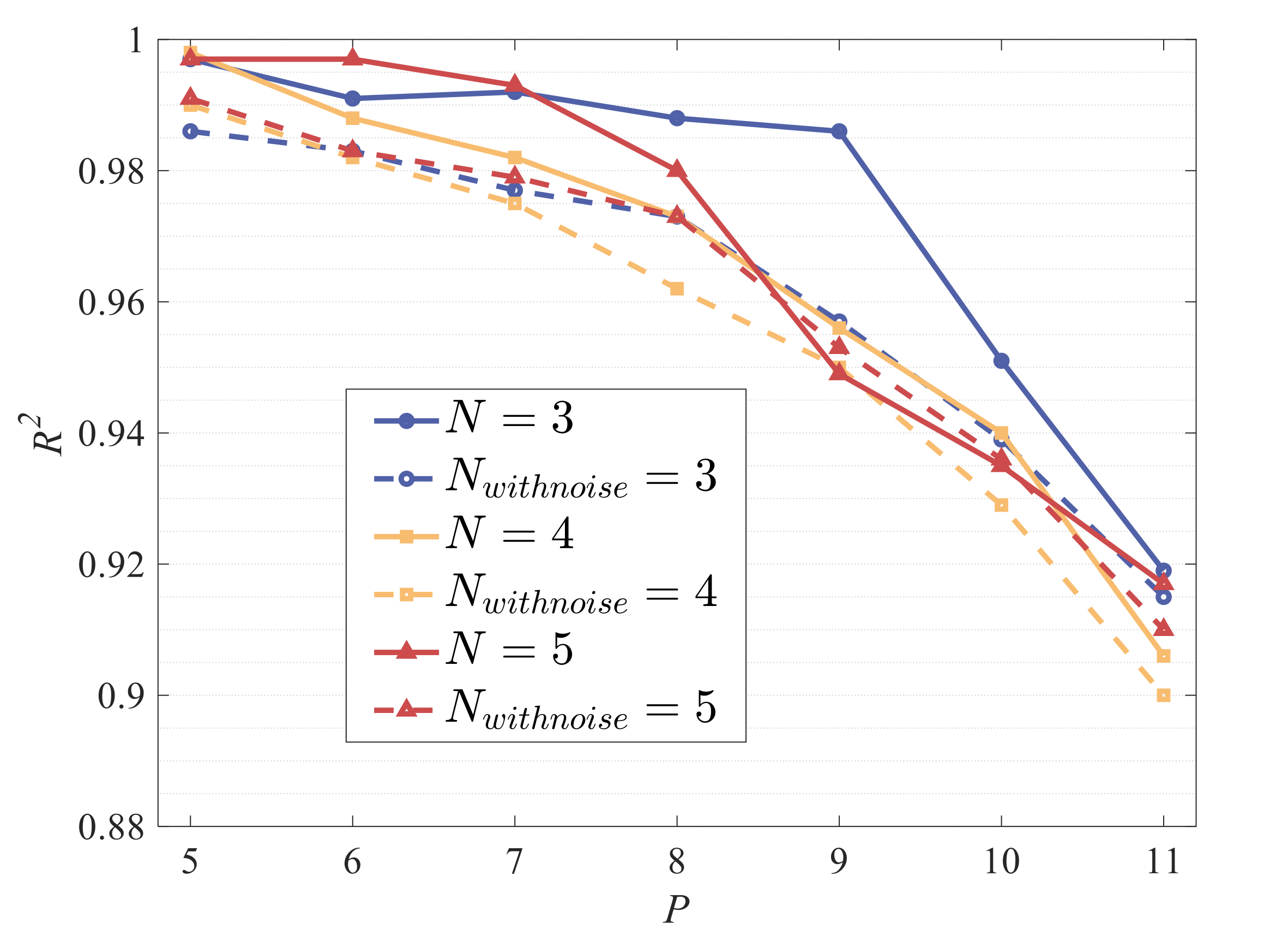}
    \captionsetup{justification=justified, singlelinecheck=false}
    \caption{Single-qubit expectation values prediction by GNNs under noisy and noiseless conditions.} \label{figure4_2a}
\end{figure}

\ 
\newline
\indent To evaluate the performance of our proposed Graph Neural Networks (GNNs)-based approach for predicting single-qubit expectation values of circuits, we randomly generate a dataset of 20000 circuits for each combination of qubit number $N$ and circuit depth $P$, where $N$ is set to 3, 4, and 5, and $P$ ranges from 5 to 11. Among these, 70$\%$ are used as the training set, 20$\%$ as the validation set, and the remaining 10$\%$ as the test set. The gate set used to construct the circuits is $S=\left \{ T, H, CNOT\right \} $. When predicting single-qubit expectation values, we focus exclusively on the first qubit, corresponding to the rescaled expectation value $z_{i}$ defined in equation (\ref{equation3_1c}), where $i=1$. It is worth noting that the same GNNs model can predict the single-qubit expectation values for any qubit by simply swapping the target qubit with the first qubit. Under noiseless conditions, the test results about $R^{2}$ values are summarized in Table \ref{table1}. Under noisy conditions, the test results are presented in Table \ref{table2}. In the noisy case, the noise model used in the simulation is derived from the IBM Perth, incorporating gate errors, $T_1$, $T_2$, and readout errors.
\begin{table}[!htp]
    \centering
    \caption{Prediction of Single-qubit Expectation Values Under Noiseless Conditions} \label{table1}
    \resizebox{0.66 \textwidth}{!}{
    \begin{tabular}{c|ccccccc} 
    \toprule
    \diagbox{$N$}{$P$} &    5 &    6 &   7 &   8 &   9 &   10 &   11 \\
    \midrule
    \midrule
    3      &    0.997   &   0.991 &   0.992 &   0.988 &   0.968 &   0.951 &   0.919 \\ 
    4  & \textbf{0.998} &   0.988 &   0.982 &   0.973 &   0.956 &   0.940 &   \textbf{0.906} \\ 
    5      &    0.997   &   0.997 &   0.993 &   0.980 &   0.949 &   0.935 &   0.917 \\ 
    \bottomrule
    \end{tabular}}
 \end{table}
\begin{table}[!htp]
    \centering
    \caption{Prediction of Single-qubit Expectation Values Under Noisy Conditions} \label{table2}
    \resizebox{0.66 \textwidth}{!}{
    \begin{tabular}{c|ccccccc} 
    \toprule
    \diagbox{$N$}{$P$} &    5 &    6 &   7 &   8 &   9 &   10 &   11 \\
    \midrule
    \midrule
    3      &    0.986 &   0.983 &   0.977 &   0.973 &   0.957 &   0.939 &   0.915 \\ 
    4      &    0.990 &   0.982 &   0.975 &   0.962 &   0.950 &   0.929 &   \textbf{0.900}   \\ 
    5      &    \textbf{0.991} &   0.983 &   0.979 &   0.973 &   0.953 &   0.936 &   0.910  \\ 
    \bottomrule
    \end{tabular}}
 \end{table}
 From Table \ref{table1}, it is evident that GNNs exhibit a significant advantage in predicting single-qubit expectation values. With a dataset of only 20000 circuits, the GNNs maintain the $R^{2}$ values above 0.9 for circuits with 3 to 5 qubits and circuit depths ranging from 5 to 11, reaching up to 0.998. Additionally, the performance of GNNs in predicting single-qubit expectation values decreases as the circuit depth $P$ increases. This is due to the rapid growth of the space of possible circuits composed of quantum gates as the circuit depth $P$ increases, while the training dataset of 20000 circuits remains relatively small compared to the vast search space. To highlight the advantage of using GNNs for predicting single-qubit expectation values affected by device noise, we compare the data from Table \ref{table1} and Table \ref{table2} and present the results in figure \ref{figure4_2a}.
Based on the data in Table \ref{table2} and figure \ref{figure4_2a}, it can be observed that when noise information from quantum devices is included in the prediction process, the GNNs still demonstrate high accuracy in predicting the single-qubit expectation values of noisy quantum circuits. This is reflected in the  $R^{2}$ values, which remain above 0.9 for circuits with 3 to 5 qubits and circuit depths ranging from 5 to 11, with a maximum value of 0.991. Compared to the noiseless case, the prediction performance of the GNNs under noisy conditions shows only a slight degradation. This indicates that the GNNs effectively learn to utilize the additional device noise information embedded in the node features during the training process, enabling more accurate predictions.

\begin{figure}[thbp]
    \centering
    \includegraphics[width= 0.88 \textwidth]{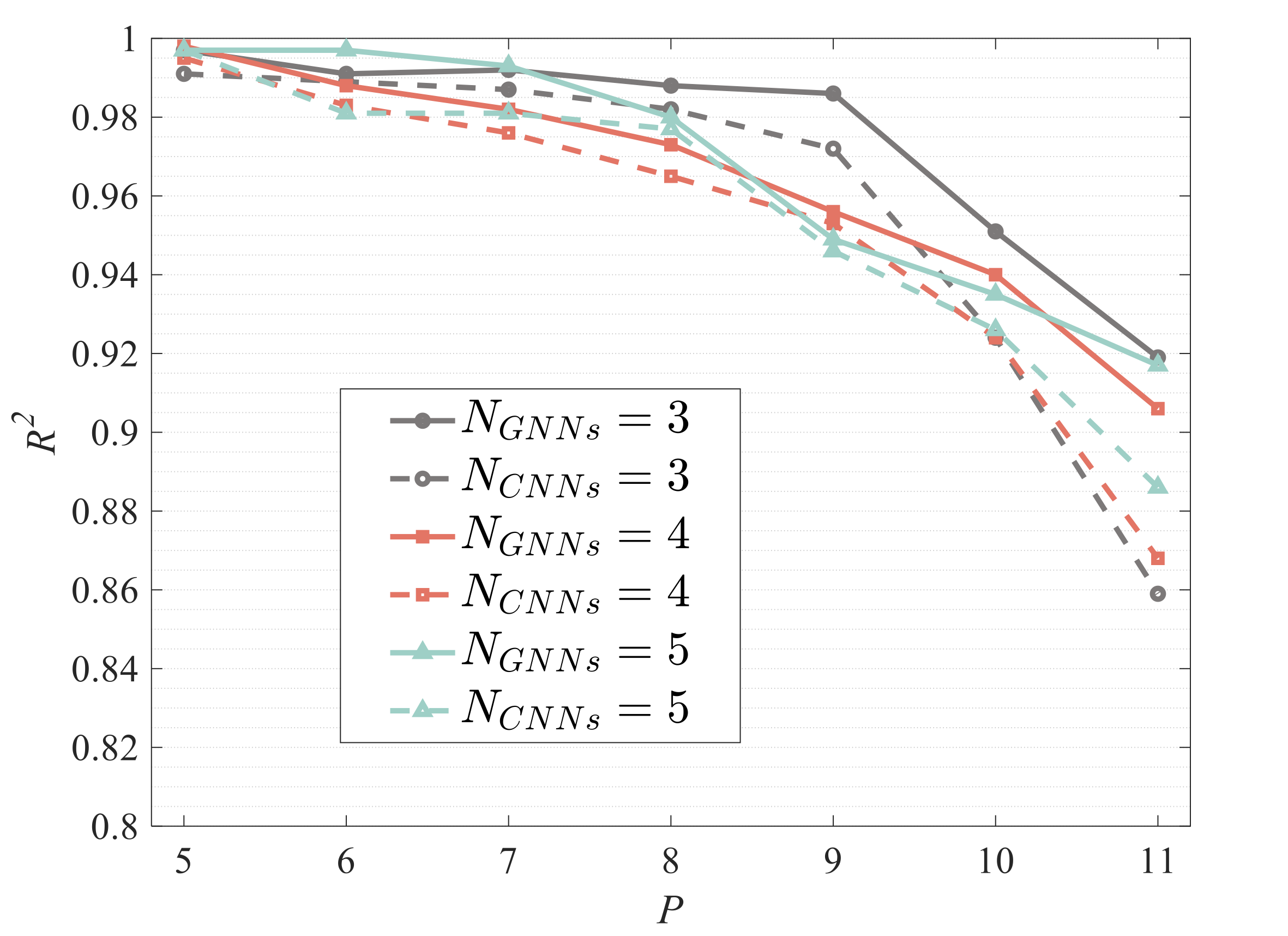}
    \captionsetup{justification=justified, singlelinecheck=false}
    \caption{Single-qubit expectation values prediction by GNNs and CNNs under noiseless conditions.} \label{figure4_2b}
\end{figure}
Furthermore, we compare our proposed method with an approach that uses Convolutional Neural Networks (CNNs) for single-qubit expectation values prediction \cite{cantori2023supervised}. The CNNs-based approach encodes quantum circuits into feature matrices using one-hot encoding, which are then fed into the networks. However, due to the requirement that CNNs need training data with identical feature dimensions, i.e., feature matrices of the same size, the CNNs-based approach cannot leverage optimization strategies for randomly generated quantum circuits with $N$ qubits and depth $P$. In addition, unlike the GNNs-based approach, which can flexibly incorporate device noise information into the feature vectors of individual nodes, the CNNs-based approach requires explicitly designed input channels to represent such noise. Considering the large variety of noise types in quantum devices and the fact that our circuits are randomly structured and vary in size, constructing a unified multi-channel representation is nontrivial. Therefore, we fully adopt the network architecture proposed in \cite{cantori2023supervised} without designing additional channels dedicated to noise representation. To evaluate and compare the performance of the two methods under noiseless conditions, we randomly generate 20000 quantum circuits using the gate set $S=\left \{ T, H, CNOT\right \} $, where the number of qubits $N$ is set to 3, 4, and 5, and $P$ ranges from 5 to 11. The single-qubit expectation values of these circuits are predicted using both the GNNs-based and CNNs-based approaches and the coefficient of determination $R^{2}$ is calculated for each. The comparative results are shown in figure \ref{figure4_2b}. As shown in figure \ref{figure4_2b}, the overall prediction accuracy$R^{2}$  decreases as the circuit depth $P$ increases. This trend is consistent with the difficulty recently reported in \cite{cantori2025challenges}, where it was observed that supervised learning becomes significantly more challenging for deeper and more highly parameterized quantum circuits. Nevertheless, figure \ref{figure4_2b} also clearly shows that the GNNs-based approach outperforms the CNNs-based approach, particularly as the circuit depth $P$ increases. This superior performance is primarily due to the fact that the topology of GNNs is derived directly from the connectivity of the quantum circuit, enabling it to better capture the interactions between quantum gates compared to the structure of CNNs. This advantage becomes increasingly pronounced as the circuit depth grows.

Unlike the CNNs-based approach, which requires the input feature matrix to have a fixed size, meaning that the training quantum circuits must have the same number of qubits $N$ and circuit depth $P$, the topology of GNNs can adapt to the structural variations of quantum circuits. This flexibility allows GNNs to be trained on quantum circuits with varying depths $P$ and even different numbers of qubits $N$. In the following, we conduct an in-depth investigation of this characteristic of GNNs.

\begin{figure*}[htbp]
    \centering
    \subfloat[]{\includegraphics[width=0.92 \textwidth]{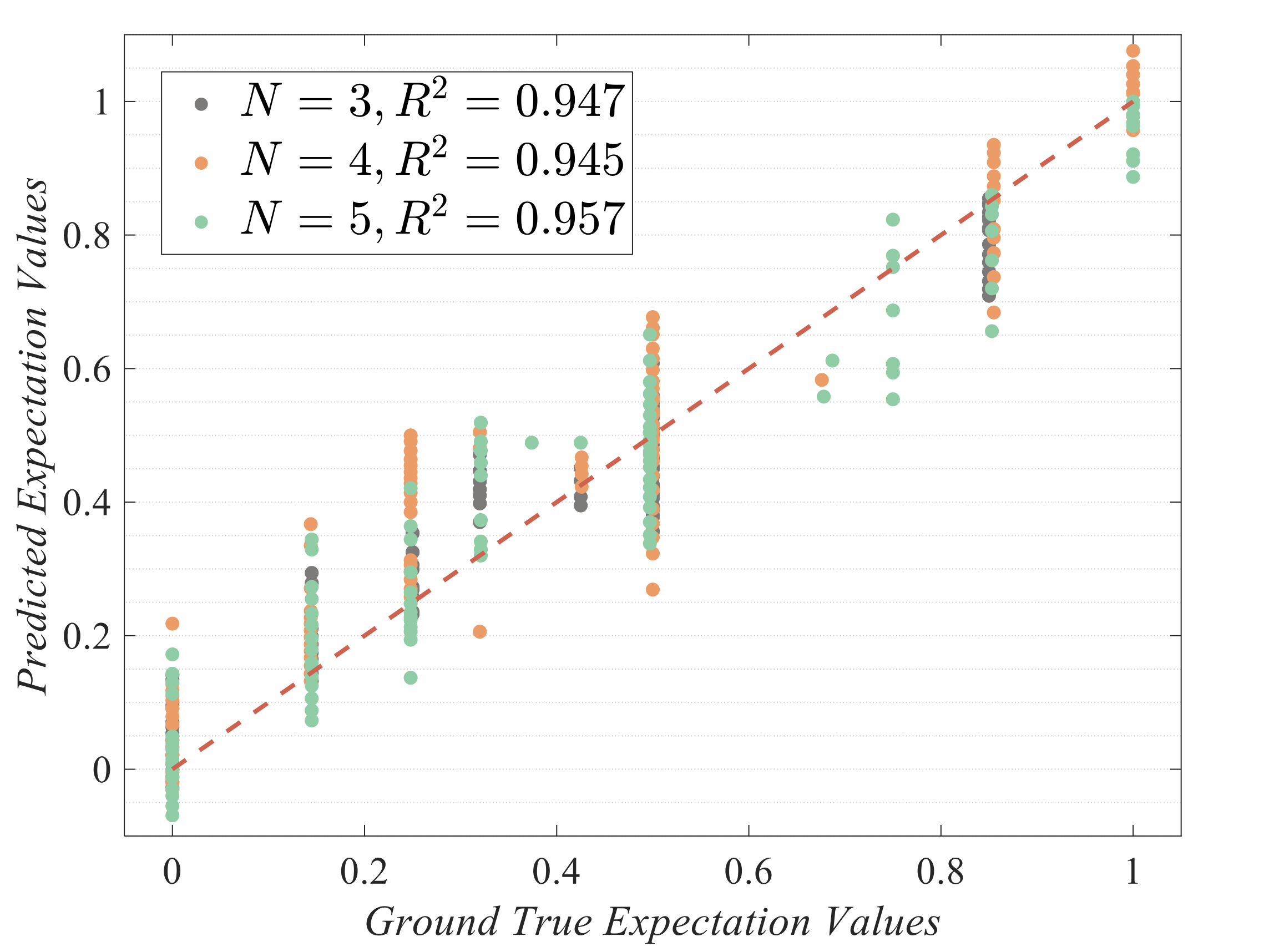}}%
    \hfil
    \subfloat[]{\includegraphics[width=0.92 \textwidth]{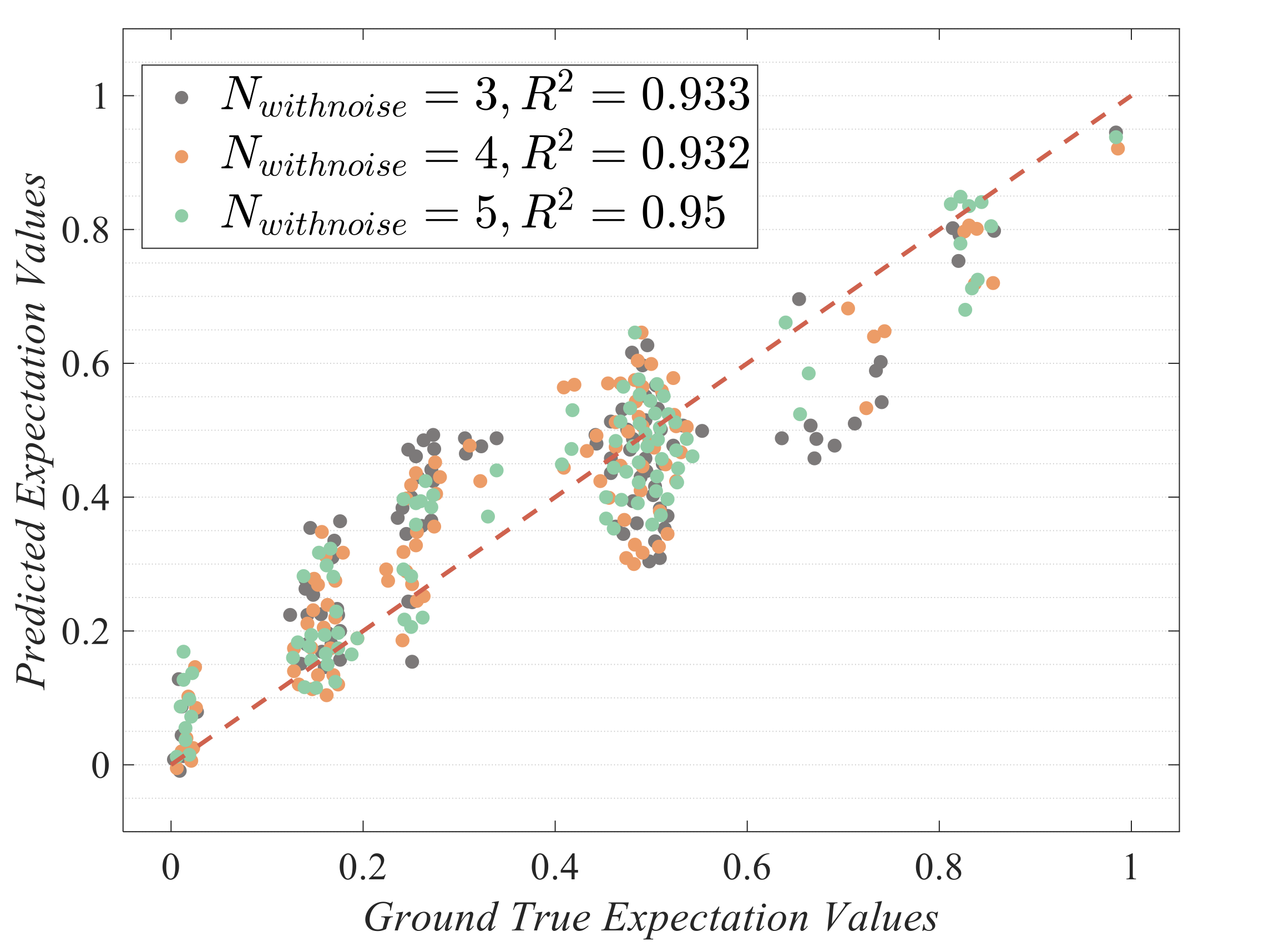}}%
    \captionsetup{justification=justified, singlelinecheck=false}
    \caption{Scatter plots of single-qubit expectation values of circuits with randomly generated depths of 5-11 under noiseless and noisy conditions.} \label{figure4_2cd}
\end{figure*}

First, we generate 20000 quantum circuits with a fixed number of qubits $N$ and circuit depths $P$ randomly distributed between 5 and 11, using the gate set $S=\left \{ T, H, CNOT\right \} $, under both noiseless and noisy conditions. The single-qubit expectation values are predicted using GNNs. Among these, 100 circuits are selected as the test set to visualize scatter plots of the predicted versus actual values and to calculate the coefficient of determination $R^2$. The prediction results under noiseless conditions are illustrated in the figure \ref{figure4_2cd}(a), while those under noisy conditions, where the noise model is derived from IBM Perth, are presented in the figure \ref{figure4_2cd}(b). The red dashed line in both panels represents the $y = x$ line.
From the scatter plot of single-qubit expectation values under noiseless conditions, we observe that the actual single-qubit expectation values primarily align with discrete values due to the gate set $S=\left \{ T, H, CNOT\right \} $ used to generate random circuits. Under noisy conditions, however, the distribution of actual single-qubit expectation values becomes more scattered as a result of noise effects. Nevertheless, since the GNNs incorporate circuit noise information during prediction, the $R^{2} $ values for predictions under noisy conditions exhibits only a negligible decline compared to the noiseless case.

\begin{figure}[thbp]
    \centering
    \includegraphics[width= 0.92 \textwidth]{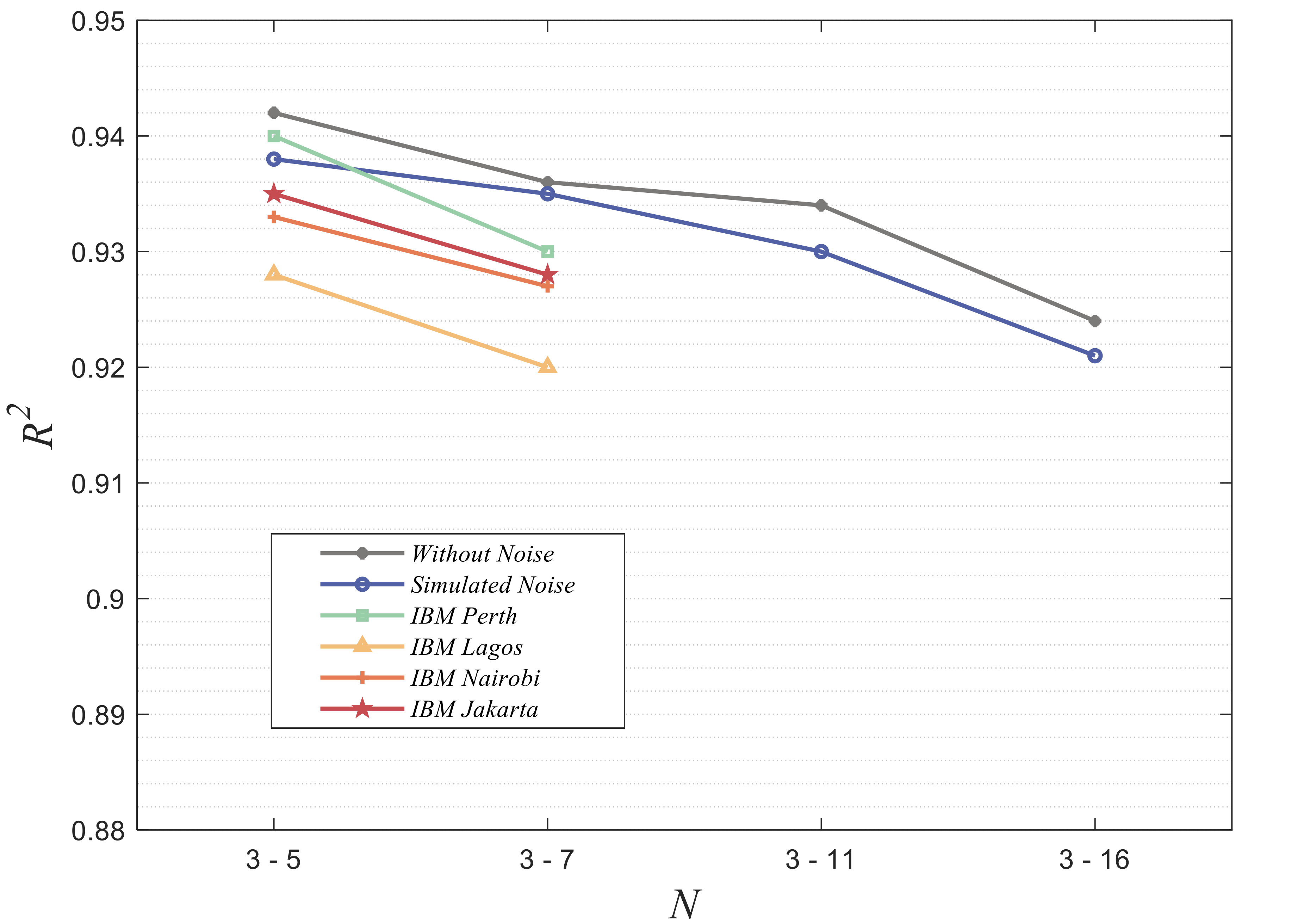}
    \captionsetup{justification=justified, singlelinecheck=false}
    \caption{Prediction of single-qubit expectation values with circuit depths of 5-11 under different conditions.} \label{figure4_2e}
\end{figure}

Subsequently, to further our investigation, we lift the restriction on a fixed number of qubits $N$ during circuit generation. Random quantum circuits are generated with different numbers of qubits $N$ and circuit depths $P$. We simulate the performance of GNNs in predicting single-qubit expectation values under noiseless conditions, Simulated Noise, and noise derived from quantum devices, including IBM Perth, IBM Lagos, IBM Nairobi, and IBM Jakarta. In all cases, 20000 random circuits with depths ranging from 5 to 11 are generated from the quantum gate set $S=\left \{ T, H, CNOT\right \} $ to construct the dataset.
Since IBM Perth, IBM Lagos, IBM Nairobi, and IBM Jakarta are 7-qubit quantum devices, simulations for these scenarios are conducted for circuits with up to 7 qubits. For noiseless and Simulated Noise scenarios, the simulations extended up to 16 qubits. Detailed comparative results about $R^{2}$ values are provided in Table \ref{table3} and illustrated in figure \ref{figure4_2e}.
\begin{table}[!htp]
    \centering
    \caption{Prediction of Single-qubit Expectation Values with Circuit Depths of 5-11 under Different Conditions} \label{table3}
    \resizebox{0.96 \textwidth}{!}{
    \begin{tabular}{c|ccccccc} 
    \toprule
    $N$      &    Without Noise & Simulated Noise & IBM Perth & IBM Lagos & IBM Nairobi & IBM Jakarta \\
    \midrule
    \midrule
    3-5      &  \textbf{0.942}  &   0.938         &   0.940    &   0.928      &   0.933     &   0.935     \\ 
    3-7      &    0.936         &   0.935         &   0.930    &\textbf{0.920} &   0.927     &   0.928     \\ 
    3-11     &    0.934         &   0.930          &   -       &   -          &   -         &   -         \\ 
    3-16     &    0.924         &   0.921         &   -       &   -          &   -         &   -         \\ 
    \bottomrule
    \end{tabular}}
 \end{table}
From Table \ref{table3} and figure \ref{figure4_2e}, we observe a slight decrease in $R^{2} $ values as the range of selectable qubits increases. However, considering the exponential growth in the potential search space of circuits brought by this expanded qubit range, this decrease is acceptable. Moreover, the GNNs achieve an $R^{2} $ value of 0.92 or higher across all conditions.

The above investigation into the characteristics of GNNs (GNNs) demonstrates that GNNs can provide reliable prediction of single-qubit expectation values for datasets containing quantum circuits with varying circuit depths and even different numbers of qubits. This significantly simplifies dataset construction, as it removes the constraint of requiring fixed input feature matrix dimensions, as is necessary for the CNNs-based approach. Furthermore, when the noise information affecting the circuits is incorporated into the feature vectors of the GNNs’ nodes, the GNNs achieve performance that is only marginally lower and sometimes nearly equivalent to that under noiseless conditions.

\subsubsection{Model extrapolation for GNNs}
\begin{figure*}[htbp]
    \centering
    \subfloat[]{\includegraphics[width=0.88 \textwidth]{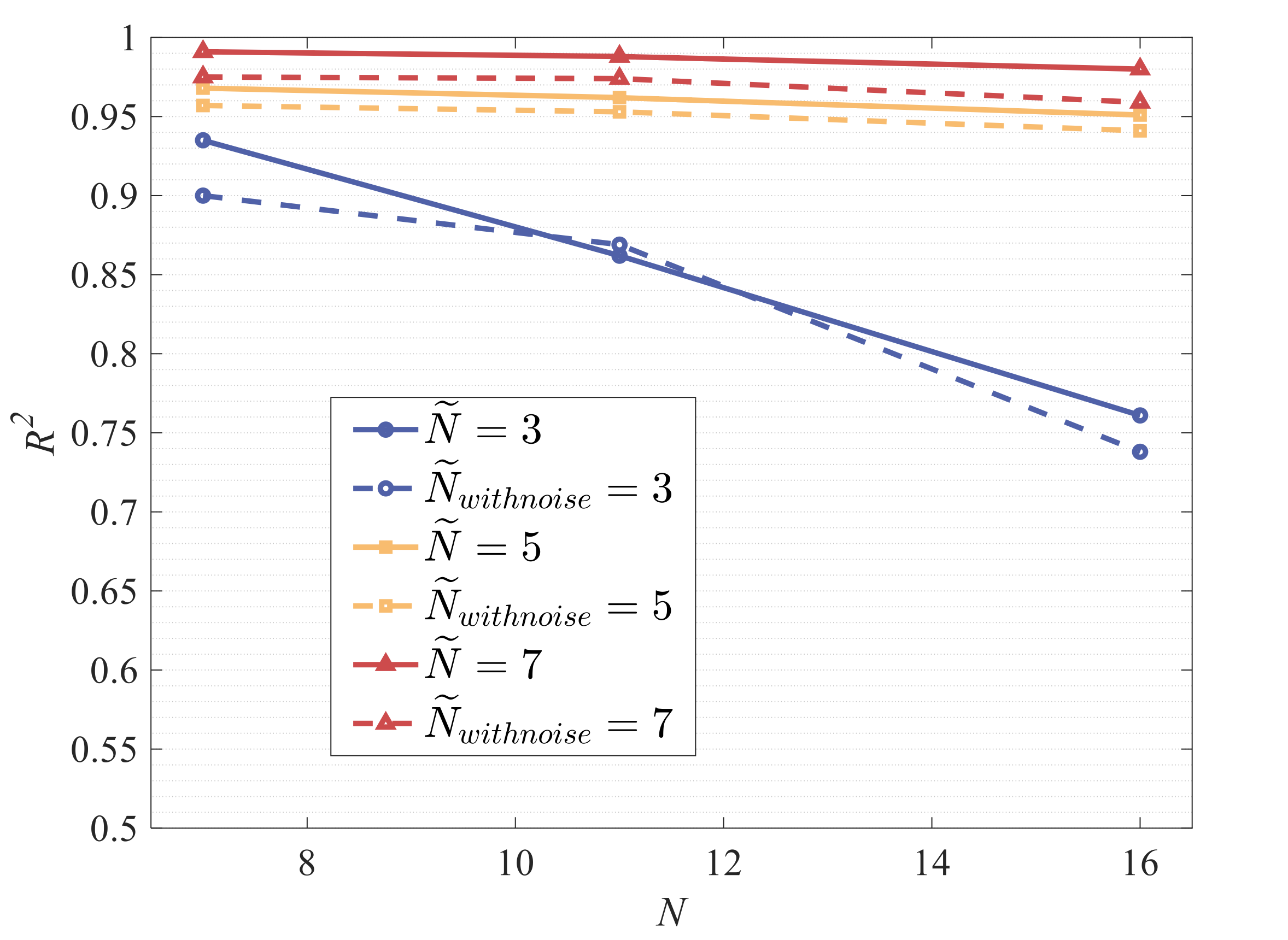}}%
    \hfil
    \subfloat[]{\includegraphics[width=0.88 \textwidth]{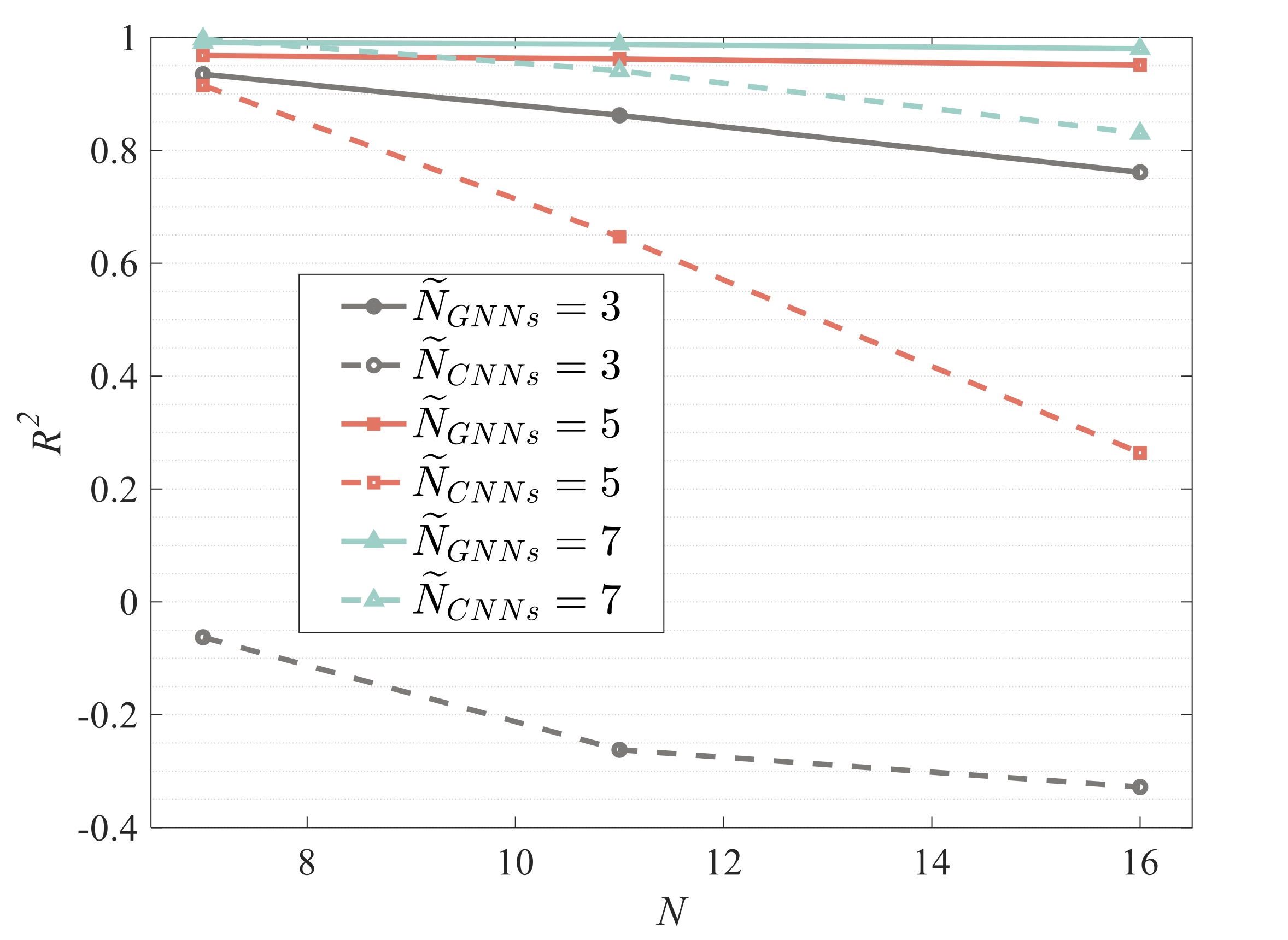}}%
    \captionsetup{justification=justified, singlelinecheck=false}
    \caption{Comparison of GNNs extrapolation performance under noiseless conditions with GNNs under noisy conditions and CNNs under noiseless conditions.} \label{figure4_2fg}
\end{figure*}

\ 
\newline
\indent Since GNNs can take quantum circuits with varying numbers of qubits as input, we partially investigate this characteristic in Section \ref{sec4.2.1}. In this subsection, we further explore the extrapolation capability of GNNs, referring to their ability to predict the properties of quantum circuits with more qubits than those in the training dataset using the same network parameters. The single-qubit expectation values are set as the prediction targets for GNNs. For each simulation, we randomly generate 40000 circuits with depths ranging from 5 to 11 from the quantum gate set $S=\left \{ T, H, CNOT\right \} $, serving as the dataset. During training, for each circuit, the ‘Gate Qubit’ part of the node feature vectors is randomly assigned the required qubits, ensuring that all qubits indicated by the ‘Gate Qubit’ part are trained. We simulate the extrapolation performance of GNNs trained on datasets of 3-qubit, 5-qubit, and 7-qubit circuits, respectively. The results comparing the extrapolation performance of these GNNs models to circuits with 7, 11, and 16 qubits under noiseless and Simulated Noise conditions are shown in the figure \ref{figure4_2fg}(a). The extrapolation performance of GNNs compared to CNNs under noiseless conditions is shown in the figure \ref{figure4_2fg}(b), and $\widetilde{N} $ represents the number of qubits in the circuits of the training set.
From the comparison in figure \ref{figure4_2fg}, it can be observed that the extrapolation performance of the GNNs model under noisy conditions is close to its performance under noiseless conditions, and both outperform the extrapolation performance of the CNNs model under noiseless conditions, particularly for 3-qubit and 5-qubit extrapolations. This advantage can plausibly be attributed to the ability of the GNNs model to better capture the connectivity between quantum gates, which facilitates improved extrapolation performance.

\subsubsection{Two-qubit expectation values}
\begin{figure}[thbp]
    \centering
    \includegraphics[width= 0.88 \textwidth]{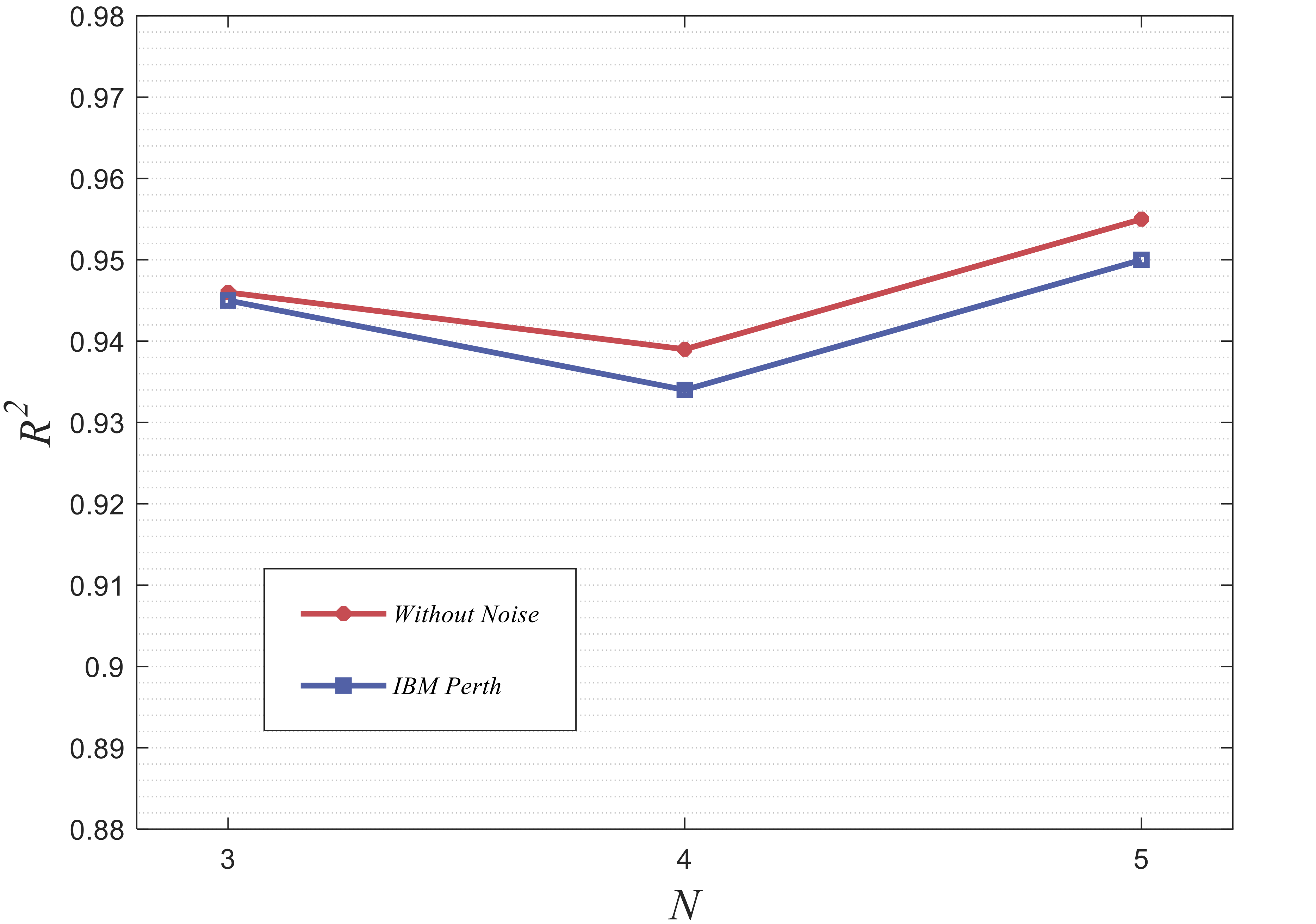}
    \captionsetup{justification=justified, singlelinecheck=false}
    \caption{Prediction of two-qubit expectation values with circuit depths of 5-11 under different conditions.} \label{figure4_2h}
\end{figure}

\ 
\newline
\indent In addition to predicting single-qubit expectation values, GNNs can also be trained to predict two-qubit expectation values. We focus on the first two qubits, corresponding to the rescaled expectation value $z_{ij}$ defined in equation (\ref{equation3_1d}). Notably, the same GNNs model can predict the two-qubit expectation value for any pair of qubits by swapping the target qubits with the first two qubits. We randomly generate 20000 circuits with depths ranging from 5 to 11 as the dataset and used GNNs to predict two-qubit expectation values under both noiseless conditions and noisy conditions with noise derived from IBM Perth. The results, including the coefficient of determination $R^{2}$, are presented in figure \ref{figure4_2h}.
Compared to the prediction of single-qubit expectation values shown in figure \ref{figure4_2cd}, GNNs demonstrate comparable performance in predicting two-qubit expectation values, achieving consistently reliable results under both noiseless and noisy conditions. This observation further motivates the exploration of broader applications for GNNs.

\subsection{Circuit performance comparison}
To extend the applicability of GNNs in PQCs, which have wide applications in the NISQ era, we investigate the capability of GNNs to predict the overall properties of a quantum circuit, specifically the ground state energy of a 4-qubit hydrogen molecule H$_{2} $. Following the approach illustrated in figure \ref{fig3_2a}, we construct PQCs using the gate set $S^{\ast } =\left \{ R_{x}\left ( \theta  \right ), R_{y}\left ( \beta  \right ), CNOT\right \} $. Subsequently, the ground state energy of H$_{2} $ is calculated using the Variational Quantum Eigensolver (VQE) algorithm, during which the parameters in the quantum circuit are iteratively optimized until convergence. The computed ground state energy of H$_{2} $ is then treated as the prediction target for GNNs and used as a performance evaluation metric for the PQCs. Based on this metric, we implement two comparative schemes for circuits with different structures. The first approach, referred to as Indirect Comparison, evaluates the performance of two quantum circuits by predicting their respective metrics separately and comparing them. We also propose a new approach called Direct Comparison, wherein GNNs directly predict the probability that one PQC outperforms another. In this scheme, the input data consists of both circuits encoded as graph structures. To obtain the labels, Min-Max normalization is first applied to the ground state energies calculated for all PQCs. The normalized ground state energy values of the two circuits are then differenced, divided by 2, and shifted by 0.5 to generate the labels for the circuit pairs. In the simulation experiments, we generate datasets of 800, 1000, 5000, and 10000 noisy PQCs, with their corresponding ground state energies calculated via the VQE algorithm. Two GNNs are trained to perform the predictions required for the Indirect Comparison and Direct Comparison schemes, respectively. 

\begin{figure}[thbp]
	\centering
	\includegraphics[width= 0.88 \textwidth]{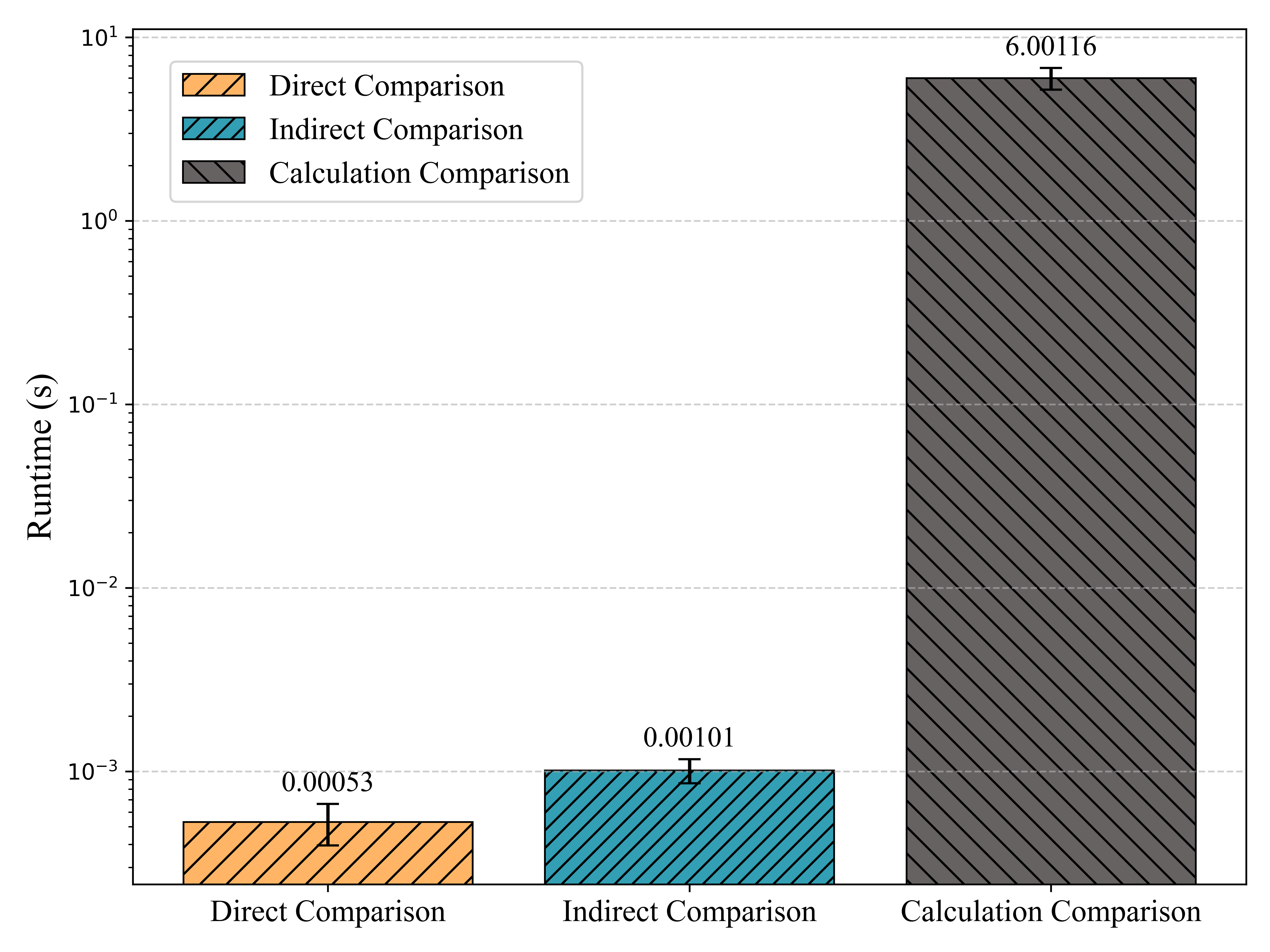}
	\captionsetup{justification=justified, singlelinecheck=false}
	\caption{Comparison of runtime of Direct Comparison, Indirect Comparison and Calculation Comparison.} \label{figure4_3add}
\end{figure}

Before comparing the prediction accuracy between the Direct and Indirect Comparison schemes, we first evaluate the performance of the same set of quantum circuits through 100 independent VQE experiments. Specifically, three evaluation methods are considered: (1) Direct Comparison, where GNNs predict the probability that one PQC outperforms another; (2) Indirect Comparison, where the performance of two circuits is compared based on their predicted metrics; and (3) Calculation Comparison, where both circuits are directly optimized using the VQE algorithm with the COBYLA optimizer for 200 iterations to obtain their ground-state energies. The mean and standard deviation of the results over 100 trials are summarized in Table \ref{table4} and illustrated in figure \ref{figure4_3add}, where the vertical axis is plotted on a logarithmic scale. As shown in Table \ref{table4}, the Direct Comparison and Indirect Comparison schemes using GNNs as predictors achieve $11321\times$ and $5941\times$ speedups, respectively, compared to the Calculation Comparison method based on direct VQE computation. These results demonstrate the potential of the trained GNNs to perform large-scale and rapid performance evaluation of PQCs, providing an efficient alternative to costly VQE simulations.
\begin{table}[!htp]
	\centering
	\caption{Runtime of Direct Comparison, Indirect Comparison and Calculation Comparison} \label{table4}
	\resizebox{0.76 \textwidth}{!}{
		\begin{tabular}{c|ccccccc} 
			\toprule
			Scheme   & Direct Comparison            & Indirect Comparison   & Calculation Comparison  \\
			\midrule
			\midrule
			Runtime $\left ( \cdot s \right )$   & $0.00053\pm 0.0001$ & $0.00101\pm 0.0002$ & $6.00012\pm 0.8178$ \\ 
			Speedup  & $\mathbf{11321\times}$ & $\mathbf{5941\times}$ & $\mathbf{1\times}$ \\ 
			\bottomrule
	\end{tabular}}
\end{table}

\begin{figure}[thbp]
    \centering
    \includegraphics[width= 0.88\textwidth]{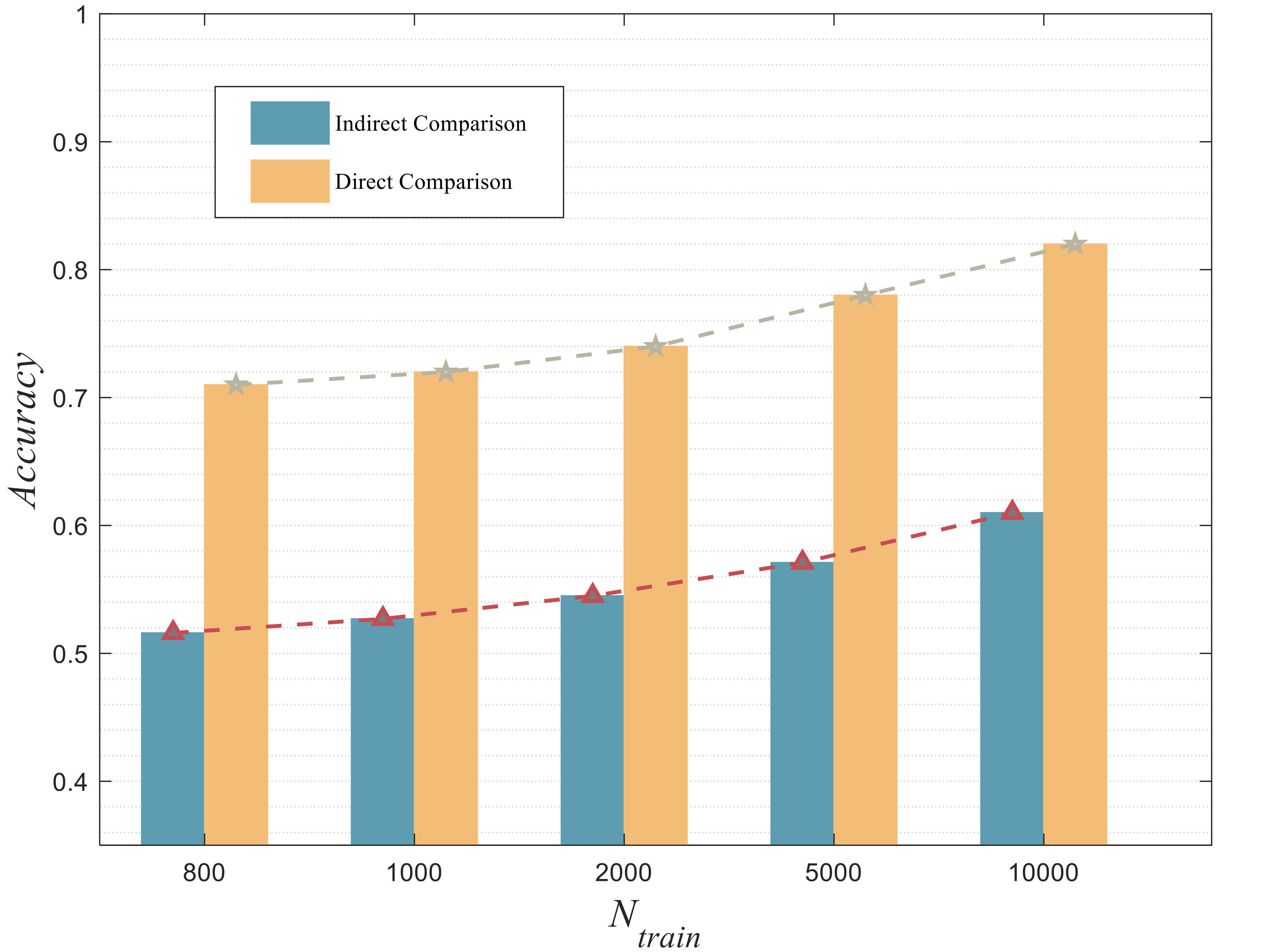}
    \captionsetup{justification=justified, singlelinecheck=false}
    \caption{Prediction accuracy of Indirect Comparison and Direct Comparison schemes for parametrized quantum circuit performance.} \label{figure4_3a}
\end{figure}

Having established the runtime advantage of the GNN-based approaches, we next assess their prediction accuracy. The prediction results are compared with the actual outcomes to evaluate the accuracy of the performance comparisons between different circuit structures. Detailed simulation results are presented in Table \ref{table5} and figure \ref{figure4_3a}, where $N_{train} $ represents the amount of data used for training.
\begin{table}[!htp]
    \centering
    \caption{Prediction Accuracy of Indirect Comparison and Direct Comparison Schemes for Parametrized Quantum Circuit Performance} \label{table5}
    \resizebox{0.76 \textwidth}{!}{
    \begin{tabular}{c|ccccccc} 
    \toprule
    \diagbox{Scheme}{$N_{train}$} & 800            & 1000  & 2000  & 5000  & 10000         \\
    \midrule
    \midrule
    Indirect Comparison           & \textbf{0.516} & 0.527 & 0.545 & 0.571 & 0.611          \\ 
    Direct Comparison             & 0.713           & 0.721  & 0.740  & 0.779  & \textbf{0.820} \\ 
    \bottomrule
    \end{tabular}}
 \end{table}
From Table \ref{table5} and figure \ref{figure4_3a}, we observe that the Direct Comparison scheme, which directly predicts the relative performance between two PQCs of different structures, significantly outperforms the Indirect Comparison scheme by an average of 36.2\% on the same dataset. While the indirect comparison scheme can enhance prediction accuracy by increasing the data volume, this approach demands substantial computational resources, particularly when executing the VQE algorithm on noisy PQCs. Additionally, if one aims to obtain a complete performance ranking of all circuits, employing a Bubble Sort algorithm can achieve this objective. The Direct Comparison scheme, in addition to providing superior circuit performance predictions over the Indirect Comparison scheme, offers an alternative perspective for the design of GNNs circuit predictors. For example, in the design of parameterized quantum circuits (PQCs), when facing a large number of candidate circuits to be screened and some high-performing reference circuits are already available, the Direct Comparison scheme can leverage these benchmark circuits to rapidly eliminate a large number of mediocre candidates, thereby reducing the cost of individually evaluating each parameterized circuit. This not only broadens the application scope of GNNs as circuit performance predictors but also contributes to the advancement of efficient PQC evaluation strategies in quantum computing.

\section{Conclusion}
\label{sec5}
We investigate the application of Graph Neural Networks (GNNs) to predict the outputs of quantum circuits, which can be constructed using either non-parameterized or parameterized quantum gates. The predictions encompass single-qubit expectation values, two-qubit expectation values, and the overall property of a quantum circuit. In our simulations, the feature vectors incorporate device noise information, including gate errors, $T_1$, $T_2$, and readout errors. As a result, the prediction performance under noisy conditions is comparable to that under noiseless conditions. Furthermore, when compared to qubit-scalable CNNs-based method, the GNNs-based approach demonstrates superior prediction performance, particularly in scenarios involving scaling from few-qubit to many-qubit circuits. In simulations comparing the Direct Comparison and Indirect Comparison schemes, the Direct Comparison scheme exhibits a significant performance improvement, establishing a robust foundation for applying GNNs-based predictors in PQCs. In the future, we aim to extend the GNNs-based predictor to predict other circuit properties using the same framework and apply it to quantum qrchitecture search (QAS) and quantum kernel design (QKD). While our simulations indicate that the predictor possesses a certain degree of extrapolation ability, how to further improve this ability is left for future investigation.

\section*{Data availability statement}
The data that support the findings of this study are openly available at the following URL/DOI: \href{https://github.com/QUANTUM-AND-ML/QaML.git}{https://github.com/QUANTUM-AND-ML/QaML.git}.

\section*{Acknowledgements}
This work is supported by the National Natural Science Foundation of China (Grant No. 62471126), the Jiangsu Frontier Technology Research and Development Plan (Grant No. BF2025066), the Fundamental Research Funds for the Central Universities (Grant No. 2242022k60001), the Jiangsu Funding Program for Excellent Postdoctoral Talent (Grant No.2022ZB139), and the Natural Science Foundation of the Higher Education Institutions of Jiangsu Province (Grant No.102024097).

\appendix
\section{Parameters of the noise model}
\label{Appendix_A}

The noise information used in our experiments for the five noise models (IBM Perth, IBM Lagos, IBM Nairobi, IBM Jakarta, and the Customized Simulated Noise) is summarized in Tables \ref{table6}–\ref{table10}. In real quantum devices, the supported gate sets vary, and different quantum gates may exhibit distinct noise characteristics depending on the qubits on which they are applied. For experimental convenience and without affecting the validity of the results, we categorize the error into single-qubit gate error and two-qubit gate error, while neglecting the effects of the actual device topology. In addition, in our Customized Simulated Noise model, all qubits share identical parameter values.

\begin{table}[!htp]
	\centering
	\caption{The Noise Parameters of the IBM Perth device} \label{table6}
	\resizebox{0.96 \textwidth}{!}{
		\begin{tabular}{c|ccccccc} 
			\toprule
			Qubit   & $T_1$ $\left ( \cdot us  \right ) $ &  $T_2$ $\left ( \cdot us  \right ) $  & Single-qubit Gate Error &  Two-qubit Gate Error & Readout Error 0-1 &  Readout Error 1-0 \\
			\midrule
			\midrule
			1           & 197.79  & 97.02  & 0.0001836  & 0.00735  & 0.0310  & 0.0226\\ 
			2           & 158.07  & 48.06  & 0.0003291  & 0.00770 & 0.0222  & 0.0236\\
			3           & 278.36  & 83.10  & 0.0002153  & 0.00694    & 0.0282  & 0.0230\\
			4           & 223.29  & 211.00 & 0.0002381  & 0.00835    & 0.0180  & 0.0130\\
			5           & 109.91  & 126.96 & 0.0003141  & 0.01169    & 0.0230  & 0.0164\\
			6           & 236.63  & 188.44 & 0.0002563  & 0.00981  & 0.0264  & 0.0208\\
			7           & 193.96  & 291.83 & 0.0003357  & 0.00984    & 0.0072  & 0.0054\\ 
			\bottomrule
	\end{tabular}}
\end{table}
\begin{table}[!htp]
	\centering
	\caption{The Noise Parameters of the IBM Lagos device} \label{table7}
	\resizebox{0.96 \textwidth}{!}{
		\begin{tabular}{c|ccccccc} 
			\toprule
			Qubit   & $T_1$ $\left ( \cdot us  \right ) $ &  $T_2$ $\left ( \cdot us  \right ) $  & Single-qubit Gate Error &  Two-qubit Gate Error & Readout Error 0-1 &  Readout Error 1-0 \\
			\midrule
			\midrule
			1           & 96.00   & 40.53  & 0.0001813  & 0.00655   & 0.0180  & 0.0092\\
			2           & 131.71  & 75.18  & 0.0001439  & 0.00770 & 0.0192  & 0.0178\\
			3           & 174.23  & 160.96 & 0.0002279  & 0.00685    & 0.0094  & 0.0074\\
			4           & 139.45  & 84.68  & 0.0001988  & 0.00651   & 0.0166  & 0.0144\\
			5           & 113.32  & 29.50  & 0.0002090  & 0.00723    & 0.0210  & 0.0198\\
			6           & 131.19  & 64.52  & 0.0001975  & 0.00685  & 0.0148  & 0.0190\\
			7           & 93.17   & 81.31  & 0.0001845  & 0.00588    & 0.0130  & 0.0152\\
			\bottomrule
	\end{tabular}}
\end{table}
\begin{table}[!htp]
	\centering
	\caption{The Noise Parameters of the IBM Nairobi device} \label{table8}
	\resizebox{0.96 \textwidth}{!}{
		\begin{tabular}{c|ccccccc} 
			\toprule
			Qubit   & $T_1$ $\left ( \cdot us  \right ) $ &  $T_2$ $\left ( \cdot us  \right ) $  & Single-qubit Gate Error &  Two-qubit Gate Error & Readout Error 0-1 &  Readout Error 1-0 \\
			\midrule
			\midrule
			1           & 104.92  & 26.07  & 0.0002782  & 0.01034   & 0.0318  & 0.0110\\
			2           & 141.98  & 97.62  & 0.0005325  & 0.01765   & 0.0720  & 0.0198\\
			3           & 57.67   & 69.22  & 0.0036500  & 0.03446   & 0.0406  & 0.0116\\
			4           & 145.77  & 59.13  & 0.0003363  & 0.01259   & 0.0366  & 0.0096\\
			5           & 99.10   & 61.06  & 0.0002725  & 0.00954   & 0.0326  & 0.0098\\
			6           & 126.61  & 16.25  & 0.0002855  & 0.01118 & 0.0432  & 0.0204\\
			7           & 148.77  & 105.21 & 0.0002028  & 0.00696   & 0.0410  & 0.0142\\
			\bottomrule
	\end{tabular}}
\end{table}
\begin{table}[!htp]
	\centering
	\caption{The Noise Parameters of the IBM Jakarta device} \label{table9}
	\resizebox{0.96 \textwidth}{!}{
		\begin{tabular}{c|ccccccc} 
			\toprule
			Qubit   & $T_1$ $\left ( \cdot us  \right ) $ &  $T_2$ $\left ( \cdot us  \right ) $  & Single-qubit Gate Error &  Two-qubit Gate Error & Readout Error 0-1 &  Readout Error 1-0 \\
			\midrule
			\midrule
			1           & 145.55  & 47.91  & 0.0003194  & 0.00806   & 0.0282  & 0.0070\\
			2           & 143.21  & 28.18  & 0.0002136  & 0.00872   & 0.0396  & 0.0096\\
			3           & 110.89  & 22.67  & 0.0002076  & 0.00914   & 0.0252  & 0.0082\\
			4           & 78.53   & 36.52  & 0.0002149  & 0.00776   & 0.0254  & 0.0112\\
			5           & 145.55  & 47.91  & 0.0003194  & 0.00806   & 0.0282  & 0.0070\\
			6           & 128.70  & 65.54  & 0.0002627  & 0.00616   & 0.0646  & 0.0480\\
			7           & 139.55  & 21.58  & 0.0002507  & 0.00577   & 0.0400  & 0.0244\\
			\bottomrule
	\end{tabular}}
\end{table}
\begin{table}[!htp]
	\centering
	\caption{The Noise Parameters of Customized Simulated Noise} \label{table10}
	\resizebox{0.96 \textwidth}{!}{
		\begin{tabular}{c|ccccccc} 
			\toprule
			Qubit   & $T_1$ $\left ( \cdot us  \right ) $ &  $T_2$ $\left ( \cdot us  \right ) $  & Single-qubit Gate Error &  Two-qubit Gate Error & Readout Error 0-1 &  Readout Error 1-0 \\
			\midrule
			\midrule
			1           & 197.79  & 97.02  & 0.0001  & 0.0083  & 0.01  & 0.01\\ 
			$\vdots $ & $\vdots $  & $\vdots $  & $\vdots $  & $\vdots $  & $\vdots $  & $\vdots $\\ 
			16           & 197.79  & 97.02  & 0.0001  & 0.0083  & 0.01  & 0.01\\ 

			\bottomrule
	\end{tabular}}
\end{table}

\section*{References}

\bibliographystyle{unsrt}
\bibliography{References}

\end{document}